\documentclass[12pt]{iopart}

\pdfoutput = 1
\usepackage{bm}
\usepackage{color}
\usepackage{graphicx}
\usepackage{citesort}
\usepackage{cancel}
\eqnobysec

\newcommand{\lambdabar}{{\hbox{$\lambda$\kern-1.ex\raise+0.45ex\hbox{--}}}}

\DeclareMathAlphabet{\mathpzc}{OT1}{pzc}{m}{it}

\begin{document}

\begin{flushright}
{\large \tt 
TTK-11-55}
\end{flushright}

\title{Evidence for dark matter modulation in CoGeNT?}

\author{Chiara~Arina\dag, Jan~Hamann\ddag, Roberto~Trotta\S\  and Yvonne~Y.~Y.~Wong\dag}

\address{\dag\  
Institut f\"ur Theoretische Teilchenphysik und Kosmologie, RWTH Aachen, 52056~Aachen, Germany}
\address{\ddag\  
Department of Physics and Astronomy, University of Aarhus, 8000 Aarhus C, Denmark}
\address{\S\  
Astrophysics Group, Imperial College London, Blackett Laboratory, Prince Consort Road, London SW7 2AZ, United Kingdom}

\eads{\mailto{chiara.arina@physik.rwth-aachen.de}, \mailto{hamann@phys.au.dk}, \mailto{r.trotta@imperial.ac.uk}, \mailto{yvonne.wong@physik.rwth-aachen.de}}

\begin{abstract}
We investigate the question of whether the recent modulation signal claimed by CoGeNT is best explained by the dark matter (DM) hypothesis from a Bayesian model comparison perspective. We consider five phenomenological explanations for the data: no modulation signal, modulation due to DM, modulation due to DM compatible with the total CoGeNT rate, and a signal coming from other physics with a free phase but annual period, or with a free phase and a free period. In each scenario, we assign to the free parameters physically motivated priors. We find that while the DM models are weakly preferred to the no modulation model, but when compared to models where the modulation is due to other physics, the DM hypothesis is favoured with odds ranging from $185:1$ to  $560:1$. This result is robust even when astrophysical uncertainties are taken into account and the impact of priors assessed.  Interestingly, the odds for the DM model in which the modulation signal is compatible with the total rate against a DM model in which this prior is not implemented is only $5:8$, in spite of the former's prediction of a modulation amplitude in the energy range $0.9 \to 3.0$~keVee that is significantly smaller than  the value observed by CoGeNT. Classical hypothesis testing also rules out the null hypothesis of no modulation at the $1.6\sigma$ to $2.3\sigma$ level, depending on the details of the alternative. Lastly, we investigate whether anisotropic velocity distributions can help to mitigate the tension between the CoGeNT total and modulated rates, and find encouraging results.

\end{abstract}

\maketitle

\section{Introduction\label{sec:intro}}
As suggested by a host of independent cosmological and astrophysical observations, most of the matter content in the Universe is invisible and reveals its existence only through gravitational effects. This is the so-called dark matter (DM). Amongst the most well-motivated candidates to explain this dark matter 
are Weakly Interacting Massive Particles (WIMPs), which have weak-scale cross-sections and masses in the GeV to TeV range.  
Many varied efforts are currently underway to search for these particles. In particular, direct detection experiments have been designed to observe WIMPs that have clustered in the Galactic halo, through their coherent scattering with nuclei in a target material. 
 
The expected WIMP scattering rates in direct searches are generally small, and the exponential decay of the WIMP recoil-spectrum mimics that of the background at low energies. However, a specific signature can be exploited in order to disentangle a DM signal from the background, namely, 
the annual modulation of the scattering rate~\cite{Drukier:1986tm,Freese:1987wu}. This effect, expected to be of the order of a few percent, is induced by the rotation of the Earth around the Sun, which periodically changes the DM flux relative to the detector, and is known to be a smoking gun signature for DM detection.

Observation of such an annual modulation signal has long been claimed by the DAMA/LIBRA experiment, a detection that has by now reached a significance of 8.9$\sigma$~\cite{Bernabei:2008yi,Bernabei:2010mq}, and can be interpreted in terms of the scattering of a light WIMP of mass $\sim10$~GeV off sodium nuclei. Interestingly, a similar light WIMP explanation appears  to explain also the  low energy excess reported by the CoGeNT experiment in 2010~\cite{Aalseth:2010vx}, as well as  the more recent CRESST results~\cite{Angloher:2011uu}. Intriguingly however, these results are at the same time excluded by a number of null searches, most notably XENON~\cite{Angle:2011th}, and upon close scrutiny, are also mildly inconsistent with one another~\cite{Chang:2010yk,Kopp:2011yr,Kelso:2011gd}. 
Adding further to the puzzle, the CoGeNT collaboration recently reported a modulating signal in the energy range $0.5 \to 3.0$~keVee at $2.8\sigma$, with a modulation amplitude of $16.6 \pm 3.8 \%$, a period of $347 \pm 29$~days, and the minimum rate falling on October~16$\pm 12$~days~\cite{Aalseth:2011wp}, somewhat at odds with the DM prediction of December~2.

The CoGeNT modulation result has sparked a new frenzy of searches for particle physics solutions that could reconcile all existing experimental findings, e.g.,~\cite{Frandsen:2011ts,Belli:2011kw,Cumberbatch:2011jp,Farina:2011pw,Schwetz:2011xm,Fox:2011px,Hooper:2011hd,Gao:2011ka,Rajaraman:2011wf,An:2011ck,DelNobile:2011je,Foot:2011pi,Boucenna:2011hy,Frandsen:2011cg,Cerdeno:2011qv,Cline:2011zr,Fornengo:2011sz,Kajiyama:2011fx,McCabe:2011sr}, as well as astrophysical solutions based on, e.g., streams~\cite{Natarajan:2011gz} and anisotropic DM velocity distributions~\cite{Frandsen:2011gi}. In this work, we examine the CoGeNT modulation data from a Bayesian model comparison perspective~\cite{Trotta:2005ar,Kunz:2006mc,Trotta:2008qt}. We ask the question of whether the data really show the presence of a modulation, and if so, whether this modulation is best explained in terms of  WIMP scattering or by some other physical process.

The classical approach of hypothesis testing does not and cannot give the probability for a hypothesis, which is not defined in frequentist statistics. Rather, it returns the probability of observing as extreme or more extreme values of the test statistic assuming the null hypothesis is true, i.e., the p-value. This, however, is in general not the scientific question one is interested in (see, e.g.,~\cite{Trotta:2008qt,Sellke:2001} for a review). In order to explain a phenomenon, one would like to assess the probability of competing models under the data. This can be done only in the context of Bayesian inference, where a probability can be assigned to all propositions---be they propositions of a parameter value within a given model (such as in the practice of Bayesian parameter inference), or of the model itself. In this way, the task of selecting the ``best'' amongst competing models to describe a set of data becomes formally one of finding the model whose posterior probability is the highest.

The central quantity in Bayesian model comparison is the evidence, or marginal likelihood, of the model, i.e., the likelihood function for the model's parameters averaged over parameter space weighted by the prior probability of the parameters. By construction the evidence automatically incorporates the notion of Occam's razor; models that fit the data well are rewarded through a favourable likelihood function; models that are unpredictive (e.g., excessively broad priors)
are penalised by the larger parameter volume over which the likelihood must be averaged. Bayesian model comparison has found widespread application in cosmological data analysis, from curvature testing~\cite{Vardanyan:2011in,Vardanyan:2011in} to inflationary model selection~\cite{Gordon:2007xm,Martin:2010hh}, amongst others~\cite{Sapone:2010uy,March:2010ex,Ichikawa:2009ir}. Its use in particle physics is less common, but see, e.g.,~\cite{Feroz:2009dv,Cabrera:2010xx,Arina:2011xu}. 

Lastly, although the main thrust of our analysis is Bayesian in nature, we also highlight along the way some of the common pitfalls in the computation of classical p-values, especially in the context of DM direct searches. We use this opportunity to recalculate the p-values for DM hypotheses that have been misestimated in previous works. 

The rest of the article is organised as follows. We introduce the statistical underpinnings of our analysis in section~\ref{sec:evid}, and describe the analysis of the CoGeNT data in section~\ref{sec:cog}. After a brief review of the theory of modulation due to WIMP scattering in section~\ref{sec:theory}, we define the models for testing in section~\ref{sec:models}. Our results are presented in section~\ref{sec:res}. Section~\ref{sec:concl} contains our conclusions.

\section{Statistical approach\label{sec:evid}}

\subsection{Bayesian inference}

Given a set of parameters $\theta$ defining a model ${\mathcal M}$, we are interested to compute their posterior probability distribution function (pdf) $p(\theta | d, {\mathcal M})$ 
via Bayes' theorem, namely,
\begin{equation}
p(\theta | d, {\mathcal M}) = \frac{{\mathcal L}(\theta | {\mathcal M}) p(\theta|{\mathcal M})}{p(d|{\mathcal M})}.
\end{equation}
Here, $d$ are the data under consideration, ${\mathcal L}(\theta | {\mathcal M}) \equiv  p(d|\theta,{\mathcal M})$ the likelihood function,
and $p(\theta|{\mathcal M})$ is the prior pdf for the parameters under the model.  The quantity $p(d|{\mathcal M})$, defined as
\begin{equation}
\label{eq:evidence}
{p}(d|\mathcal{M}) \equiv \int \mathcal{L}(\theta| \mathcal{M})p(\theta|\mathcal{M}) \rmd \theta\,,
\end{equation}
is called the Bayesian evidence.

Bayesian inference is based on the posterior pdf for the parameters $\theta$, and it assumes that the model under consideration, $\mathcal M$, is the correct one. But we can expand our inferential framework to ask the question of the viability of the model itself, or rather, of the relative performance of alternative possible models as explanations for the data ---this is the subject of Bayesian model comparison. The formalism of Bayesian model comparison automatically balances the quality of the model's  fit to the data against its predictiveness. Thus, if there are several competing models, the problem of finding the ``best'' model---one that achieves the optimum compromise between quality of fit and predictiveness---can be formally defined as selecting the model that has the highest posterior probability. In this sense, the methodology of  Bayesian model selection can be interpreted as  a quantitative expression of the Occam's razor principle of simplicity. 

We now turn to a brief review of the framework of Bayesian model comparison.  For a more in-depth discussion see, e.g.~\cite{Kunz:2006mc,Trotta:2008qt}.

\subsection{Bayes factors and model comparison \label{sec:evi1}}

From the definition of the Bayesian evidence in equation~(\ref{eq:evidence}), it is easy to see that this quantity incorporates the notion of Occam's razor and penalises those models with excessive complexity unsupported by the data for wasted parameter space. Increasing the dimensionality of the parameter space without significantly enhancing the likelihood $\mathcal{L}(d|\theta, \mathcal{M})$ in the new parameter directions reduces the evidence. Unpredictive priors $p(\theta|\mathcal{M})$ (e.g., excessively broad compared with the width of the likelihood) likewise dilute the evidence.

The posterior probability $p(\mathcal{M}|d)$ of a model $\mathcal{M}$ given the data $d$ is related to the Bayesian evidence via Bayes' theorem,
\begin{equation}
p(\mathcal{M}|d) \propto p(d|\mathcal{M})\  p(\mathcal{M}) \,,
\end{equation}
where $p(\mathcal{M})$ is the prior probability assigned to the model $\mathcal{M}$ itself, and we have dropped a normalisation constant corresponding to the probability of the data $d$. Thus, the posterior odds between two competing models $\mathcal{M}_0$ and $\mathcal{M}_1$ are given by
\begin{equation}
\frac{p(\mathcal{M}_1|d)}{p(\mathcal{M}_0|d)} = B_{10}\frac{p(\mathcal{M}_1) }{p(\mathcal{M}_0) }\,,
\end{equation}
where  
\begin{equation}
\label{eq:bayesfactor}
B_{10} \equiv \frac{p(d|\mathcal{M}_1)}{p(d|\mathcal{M}_0)}\,,
\end{equation}
a ratio of the models' evidences, is called the Bayes factor.

The Bayes factor $B_{10}$ represents an update from our prior belief in the odds of two competing models $p(\mathcal{M}_1)/p(\mathcal{M}_0)$
to the posterior odds $p(\mathcal{M}_1|d)/p(\mathcal{M}_0|d)$.  If there are no prior reasons to believe that one model should be more probable than the other,
then $p(\mathcal{M}_1)=p(\mathcal{M}_0)$ (non-committal prior), and the Bayes factor alone determines the outcome of the model comparison.
A Bayes factor larger than unity means that the model $\mathcal{M}_1$ is preferred over the model $\mathcal{M}_0$ as a description of the 
experimental data, and vice versa.   As may be expected, the correspondence between the actual value of the Bayes factor and the degree of preference 
in the ordinary sense---and hence the use of the Bayes factor as a decision-making criterion---is a matter of 
convention.\footnote{In the same way, the classical statistics criterion that a null hypothesis should be rejected if the p-value falls below
 0.05 is a matter of convention.} Here, we use the convention set down by ``Jeffreys' scale'' shown in table~\ref{tab:jef}.

\begin{table}[t!]
\caption{Jeffreys' scale for grading the strength of evidence for two competing models $\mathcal{M}_0$ and $\mathcal{M}_1$, adapted from~\cite{Gordon:2007xm,Trotta:2008qt}.\label{tab:jef}}
\begin{center}
\lineup
\begin{tabular}{lll}
\br
$\ln B_{10}$ & Odds $\mathcal{M}_1: \mathcal{M}_0$& Strength of evidence \\
\mr
$<-5.0$ & $< 1:150$ & Strong evidence for $\mathcal{M}_0$ \\
$-5.0 \to -2.5$  & $1:150 \to 1:12$ & Moderate evidence for $\mathcal{M}_0$ \\
$-2.5 \to -1.0$ & $1:12 \to 1:3$ & Weak evidence for $\mathcal{M}_0 $ \\
$-1.0 \to 1.0$ & $1:3 \to 3:1$ & Inconclusive\\
$1.0 \to 2.5$ & $3:1 \to 12:1$ & Weak evidence against $\mathcal{M}_0 $ \\
$2.5 \to 5.0$  & $12:1 \to 150:1$ & Moderate evidence against $\mathcal{M}_0$ \\
$> 5.0$ &$> 150:1$ & Strong evidence against $\mathcal{M}_0$ \\
\br
\end{tabular}
\end{center}
\end{table}

As can be seen in equation~(\ref{eq:evidence}), the computation of the evidence $p(d|\mathcal{M})$ for each model $\mathcal{M}$ requires the evaluation of an integral over the parameter space. For this purpose we use the multimodal nested-sampling algorithm implemented in the publicly available package \texttt{MultiNest} v2.12~\cite{Feroz:2007kg,Feroz:2008xx}.  We set $n_{\rm live} = 10000$ and a tolerance factor of 0.01, following the recommendations of~\cite{Feroz:2007kg}.

\subsection{Sensitivity analysis for the Bayes factors} 

The evidence and the Bayes factors are obviously dependent on the choice of the prior probability distributions for the model's parameters,
$p(\theta|\mathcal{M})$. Since the choice of priors is usually not unique, interpretation of the results of Bayesian model selection ought to allow for the impact of a reasonable change of priors. This is called ``sensitivity analysis''.

As we shall describe in detail in section~\ref{sec:models}, our approach to studying time modulation in the CoGeNT data is purely phenomenological.  We begin with the 
parameter space of the most complex model (in terms of the number of free parameters it contains), $\theta = \{U_{\rm m}^i, S_{\rm m}^i, t_{\rm max},
T\}$, and define from it a set of nested models, where simpler models are realised by setting some or all of these parameters to specific
values~$\theta^\star$.  With one exception (model~1b, see table~\ref{tab:resume}), the models we
test are not tightly connected to the predictions of any specific physical process.
This approach, while perfectly convenient for the purpose of parameter inference, poses a problem for model comparison:  without the guidance
of specific predictions, the odds for a more complex model can be made arbitrarily small by increasing the width of the priors on the
additional parameters or by choosing uniform priors on non-linear functions of these parameters.

If the models $\mathcal{M}_0$ and $\mathcal{M}_1$ are nested and their parameter priors separable, then the impact of changing the prior width 
 on the Bayes factor can be estimated analytically using the Savage-Dickey density ratio (SDDR, see~\cite{Trotta:2005ar}), 
\begin{equation}
\label{eq:sddr}
B_{10} = \frac{p(\vartheta^{\star}|\mathcal{M}_1)}{p(\vartheta^{\star}|d,\mathcal{M}_1)}\,.
\end{equation} 
Introducing the notation $\theta = (\vartheta,\psi)$, where $\vartheta$ denotes the $N$ additional parameters of the more complex model $\mathcal{M}_1$, and $\psi$ labels the free parameters common to both models, $p(\vartheta^{\star}|d,\mathcal{M}_1)
\equiv \int p(\vartheta^{\star},\psi |d,\mathcal{M}_1) \rmd \psi$ is
the marginal posterior pdf for the additional parameters of $\mathcal{M}_1$ evaluated at $\mathcal{M}_0$'s parameter value $\vartheta^\star$, marginalised over~$\psi$, and $p(\vartheta^{\star}|\mathcal{M}_1) \equiv \int p(\vartheta^{\star},\psi |\mathcal{M}_1) \rmd \psi$ is the marginal prior density of $\mathcal{M}_1$ evaluated at the same $\vartheta^\star$.

If the prior $p(\vartheta|\mathcal{M}_1)$ is sufficiently broad and the data sufficiently constraining, then $p(\vartheta^{\star}|d,\mathcal{M}_1)$ will be approximately
independent of the prior.  However, because the prior pdf must be normalised to unity probability content, increasing the width of $p(\vartheta|\mathcal{M}_1)$ leads to a smaller $p(\vartheta^{\star}|\mathcal{M}_1)$ and therefore to a smaller Bayes factor.  If for instance $p(\vartheta |\mathcal{M}_1)$ is a top-hat function, the SDDR formula shows that rescaling its width by a factor $\alpha$ will change $\ln B_{10}$ by approximately $-\ln \alpha$. We will use this analytic approximation to perform a sensitivity analysis of our model comparison results.

\section{CoGeNT data analysis\label{sec:cog}}

The CoGeNT  collaboration has recently released  data corresponding to 458~days of data taking, of which 442~days are live~\cite{Aalseth:2011wp}. The initial day $t_{\rm in}$ is December~4, 2009, from which point on data taking has been continuous until March~6, 2011,
except for the gaps on days $68 \to 74$, $102 \to 107$, and $306 \to 308$ inclusive. These gaps are taken into account  in our time-binning of the data. The fiducial detector mass is 330~g, leading to a total exposure of 145.86~kg-days. Although the energy threshold of the detector is nominally 0.4~keVee (electron equivalent keV),  we consider in this work a threshold of 0.5~keVee in order to avoid trigger threshold effects. 

The signal region receives contributions from events induced by cosmogenic activation of radioisotope decays via electron capture. Several peaks are expected at energies predicted by $K$-shell decay ranging from 0.56 to 1.4~keVee, the dominant being the Ge$^{68}$ and the Zn$^{65}$ peaks at 1.1 and 1.3~keVee, respectively.  
These can be subtracted following the prescription of the CoGeNT collaboration~\cite{collar}. Henceforth, we shall present event rates only {\it after} subtracting these contributions.
 
Our main focus in this work is the time modulation of the CoGeNT data. It is on the measured event rate as a function of time that we perform our Bayesian model comparison analysis. However, as we shall see in section~\ref{sec:models}, the total number of events summed over the whole measurement period can be used to construct self-consistent priors on the modulation amplitude. We describe the analysis of both the time-dependent event rate and the total event rate below.

\subsection{Time-dependent rate}

\begin{figure}[t!]
\begin{minipage}[t]{0.32\textwidth}
\centering
\includegraphics[width=1.\columnwidth]{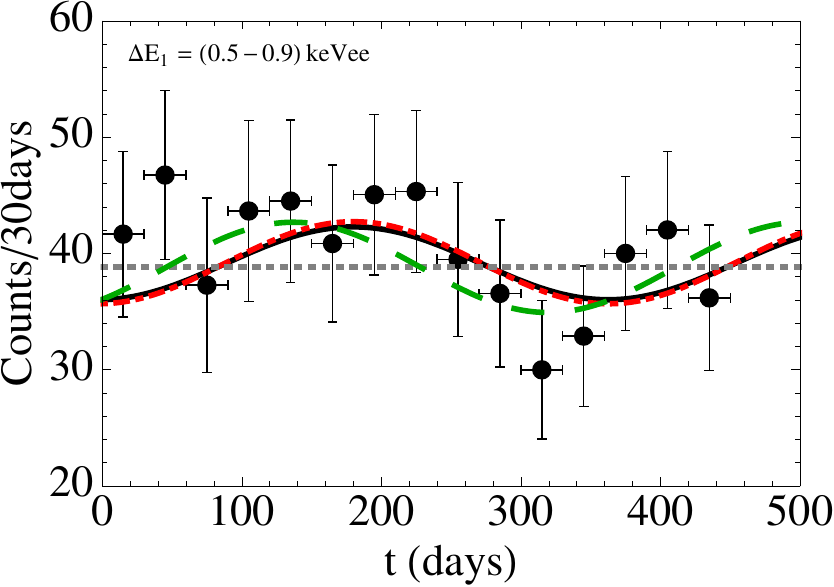}
\end{minipage}
\begin{minipage}[t]{0.32\textwidth}
\centering
\includegraphics[width=1.02\columnwidth]{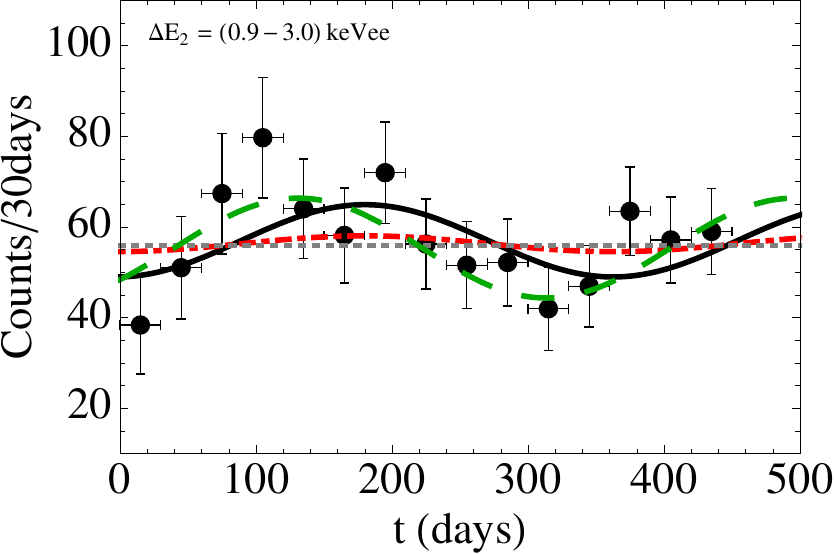}
\end{minipage}
\begin{minipage}[t]{0.32\textwidth}
\centering
\includegraphics[width=.99\columnwidth]{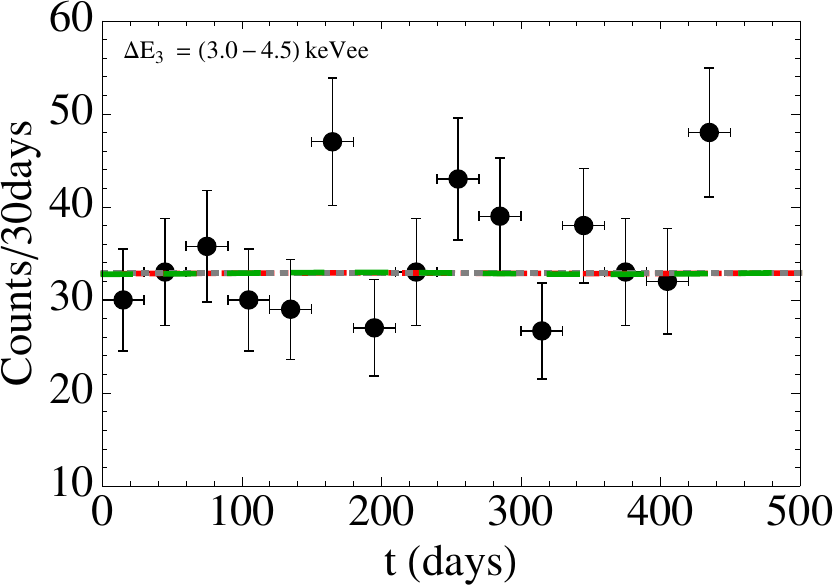}
\end{minipage}
\caption{CoGeNT time-stamped data binned in intervals of 30~days, in three energy bins.  From left to right, $\Delta E_1 = 0.5 \to 0.9$~keVee, $\Delta E_2 = 0.9 \to 3.0$~keVee, and $\Delta E_3 =3.0 \to 4.5$~keVee.  Note that contributions from radiative peaks have been removed. In each panel, the black curve denotes the best-fit point for model~1a, the red dot-dashed line for~1b, the green long-dashed line for~2a, while the dotted gray line indicates
 the constant component of the event rate.
\label{fig:CoGSmt}}
\end{figure}

We investigate the time evolution of the event rate in three top-hat energy bins (index~$i$): $\Delta E_1 =0.5 \to 0.9$~keVee, $\Delta E_2 = 0.9 \to 3.0$~keVee, and $\Delta E_3 =3.0 \to 4.5$~keVee. The choice of these bins is based on theoretical expectations for a light-mass WIMP interacting in a CoGeNT-like detector:  there should be a time modulation in the signals in $i=1,2$, and none in $i$=3.  For each energy bin, we group the time-stamped data in intervals of 30~days starting from $t_{\rm in}$. The last interval of 8~days is discarded from the analysis, because of its minor statistical weight due to the large error bar. This gives a total of 15 time bins (index $j$). We show in figure~\ref{fig:CoGSmt} the time-binned data $C_{ij}$ for the three energy bins, in units of  counts per 30~days. The corresponding error bars $\sigma_{ij}$ are estimated from Poisson statistics, i.e., $\sigma_{ij} =(30/x_j) \sqrt{\tilde{C}_{ij}}$, where $\tilde{C}_{ij}$ is the raw number of events prior to subtraction of the radiative peaks in the $i$th energy bin and the $j$th 30-day interval, and $x_j$ is the actual number of days of data taking in the $j$th time interval, in the event of gaps.

The likelihood function for the $i$th energy bin is approximated as an uncorrelated multivariate Gaussian function, i.e., up to irrelevant normalisation constants,
\begin{equation}
\ln\mathcal{L}^i_{\rm time} = -\sum_{j=1}^{15}\frac{(R_{ij} -C_{ij})^2}{2 \sigma_{ij}^2}\,,
\end{equation}
where $R_{ij}$ denotes the theoretical event rate in the $i$th energy bin and $j$th time bin.  In the case of WIMP scattering, $R_{ij}$ is a function of the DM parameters as well as astrophysical quantities. We discuss in more detail the computation of $R_{ij}$ in section~\ref{sec:models}.

\subsection{Total rate\label{sec:totalrate}}

We do not use the CoGeNT total event rate directly in our model comparison analysis. However, as we shall explain in section~\ref{sec:models}, in some cases we are interested to estimate the DM parameters preferred by the total rate, and use this information to construct self-consistent priors on the modulation amplitude to be used in  the model comparison analysis of  the time-dependent data.

To this end, we group the total number of events observed in the energy range $0.5 \to 3.2$~keVee into 27 energy bins of width $\Delta_{b} = 0.1$~keVee. In each of these bins, the quantity $C_k$ denotes the number of counts per day per detector mass per energy (in units cpd/kg/keVee) over the period of data taking, with the corresponding statistical error $\sigma_k$  estimated from Poisson statistics prior to subtraction of  the radiative peaks. The likelihood function for the total rate is approximated as a Gaussian, i.e.,
\begin{equation}
\ln\mathcal{L}_{\rm total}  = -\sum_{k=1}^{27}\frac{((S_k+b_k) - C_k)^2}{2 \sigma_k^2}\,,
\end{equation}
 where $S_k$ is the predicted DM signal rate in the $k$th energy bin, while $b_k$ is the background contribution to that bin. Note that by approximating the likelihood function as an uncorrelated multivariate Gaussian, we have implicitly assumed the count rates in the individual energy bins to be uncorrelated. This should be a reasonable assumption, given that the CoGeNT energy resolution, $\sigma_\mathcal{E} \simeq \sqrt{69.7^2 + 981 \mathcal{E}/{\rm keVee}}$~keVee~\cite{Aalseth:2008rx}, i.e.,  $\sigma_\mathcal{E} \simeq 0.02 \to 0.057$~keVee in the energy range considered in this work, is always smaller than the energy bin width $\Delta_{b}$.

The DM signal $S_k(m_{\rm DM},\sigma_n^{\rm SI}; \rho_\odot, f(\vec{v'}))$ comes from integrating the differential recoil rate for coherent and elastic WIMP scattering off nuclei over the energy range of the $k$th bin, and is a function of the DM mass $m_{\rm DM}$ and the spin-independent cross-section $\sigma_n^{\rm SI}$, as well as of the local DM density $\rho_\odot$ and several other astrophysical parameters characterising the DM velocity distribution $f(\vec{v'})$ in the solar neighbourhood. We shall elaborate more on the astrophysics in section~\ref{sec:models} where appropriate. For details of the computation of $S_k$, we refer the reader to~\cite{Arina:2011si}. Note however a technical detail that differs in the present analysis: the form of the (energy-dependent) quenching factor $q_{\rm Ge}$, which relates the observed ionisation energy
$\mathcal{E}$ (in units keVee) to the actual recoil energy $E$ (in units of nuclear recoil keV, keVnr) via $\mathcal{E}=q _{\rm Ge} E$.
Here, we use Lindhard's theory with $\kappa = 0.2$, i.e., 
\begin{equation}
\mathcal{E} ({\rm keVee}) = 0.19935 \times E^{1.1204} ({\rm keVnr}) \,,
\end{equation}
as recommended in~\cite{collar}.

\begin{table}[t!]
\caption{Priors for the nuisance parameters describing the background in the CoGeNT experiment. All priors are uniform over the indicated range.\label{tab:priorsback}}
\begin{center}
\lineup
\begin{tabular}{cl}
\br
Parameter & Prior \\
\mr
$\mathcal{E}_0$  & $0  \to 30$ (keVee)\\
 $\mathcal{C}$ & $0 \to 10 \ ({\rm cpd/kg/keVee})$\\ 
  $\mathcal{A}$ &  $0 \to 10\ ({\rm cpd/kg/keVee})$\\
\br
\end{tabular}
\end{center}
\end{table}

The background rate $b_k$ is obtained similarly from integrating over the $k$th bin the differential background $\rmd B/\rmd \mathcal{E}$, which we model as a constant factor plus an exponential decay,
\begin{equation}
\frac{{\rm d}B}{{\rm d}\mathcal{E}} = \mathcal{C} + \mathcal{A} \exp(-\mathcal{E}/\mathcal{E}_0)  \,,
\end{equation}
so that
\begin{eqnarray}
\label{eq:bckg}
b_k & = & \frac{1}{\Delta_b} \int_{\mathcal{E}_k}^{\mathcal{E}_{k}+\Delta_b} \frac{{\rm d}B}{{\rm d}\mathcal{E}} {\rm d}\mathcal{E}\,,\nonumber\\
 & = &\mathcal{C} + \frac{\mathcal{A} \mathcal{E}_0}{\Delta_b} \left[\exp\left(-\frac{\mathcal{E}_k}{\mathcal{E}_0}\right)-\exp\left(-\frac{\mathcal{E}_{k}+\Delta_b}{\mathcal{E}_0}\right)\right]\,,
\end{eqnarray}
with nuisance parameters $\mathcal C$, $\mathcal A$ and $\mathcal{E}_0$. The exponential background has been introduced to account for surface event contamination in the signal region due to misidentified electrons in the nuclear recoil band. This and the flat background rate $\mathcal C$ are assumed to be constant in time. Table~\ref{tab:priorsback} shows the prior ranges for these nuisance parameters.

\section{Annual modulation due to dark matter\label{sec:theory}}

Direct dark matter searches are sensitive to both the particle physics of the DM candidate and  the local properties of the Galactic halo in the solar neighbourhood, $\vec{R}_\odot$. The event rate due to scattering of WIMPs in a dark matter detector is proportional to 
\begin{equation}
\label{eq:fv}
\eta(E,t) =  \int_{v'>v_{\rm min}} \rmd^3v'   \,  \frac{ f (\vec{v'}(t))}{v'}\,,
\end{equation} 
where $f(\vec{v'}(t))$ is the local velocity distribution of the DM particles with respect to the Earth's frame (primed), and $v_{\rm min}$ is the minimum 
velocity required of the particles to deposit an energy $E$ in the detector. See for further details~\cite{Arina:2011si} and references therein. 

To compute $f (\vec{v'}(t))$, we note that on the timescale of a direct detection experiment, the local DM velocity distribution in the Galactic frame (unprimed),  
 $F(\vec{v}, \vec{R}_\odot)$, can be taken to be constant in time. However, because the solar system moves through the DM halo with a velocity $\vec{v}_\odot$  in the Galactic frame, and the Earth moves around the Sun with a rotation velocity $\vec{v''}_{\oplus,{\rm rot}}$ in the Sun's rest frame, the local DM velocity distribution as seen on Earth, $f (\vec{v'}(t))$, becomes time-dependent via the transformations
 \begin{eqnarray}
 \label{eq:voplus}
 v^2 &=& | \vec{v'} + \vec{v}_\oplus |^2, \nonumber \\
v_{\oplus} & = & |\vec{v}_\odot+\vec{v^{''}}_{\oplus,{\rm rot}} | = v_\odot  + v''_{\oplus,{\rm rot}}  \cos\gamma\cos [2 \pi (t-t_0)/T]\,,
\end{eqnarray}
where $\gamma=60^{\circ}$ is the inclination of the Earth's rotation plane with respect to the Galactic plane, $t_0 \sim$~June~2, and $T = 1$~year.  Note that the second equation is a projection 
of the Earth's orbital motion onto the Galactic plane and neglects the small ellipticity of the orbit.

We can easily estimate the effect of the Earth's movement on the DM scattering rate in the $|\vec{v'}+\vec{v}_\odot| \ll v^{''}_{\oplus,{\rm rot}}$ limit by expanding $\eta(E,t)$ around its ``mean'' value $\eta_0(E,t)$ (i.e., evaluated at $v^{''}_{\oplus,{\rm rot}}=0$) in terms of the small parameter $\epsilon=v^{''}_{\oplus,{\rm rot}}/|\vec{v'}+\vec{v}_\odot|$. For an isotropic halo, i.e., $F(\vec{v}, \vec{R})= F(v, R)$, we find
\begin{equation}
\label{eq:vdep}
\eta(E,t) - \eta_0(E,t) \propto \cos[2 \pi (t-t_0)/T]\,,
\end{equation}
which is the well-known result that the WIMP scattering rate has a sinusoidal time dependence with a period of one year, and a phase 
determined by the movement of the Earth with respect to the Galactic frame, $\vec{v}_{\oplus}$. We shall deal only with isotropic DM velocity distributions in our model comparison analysis, i.e., equation~(\ref{eq:vdep}) is always taken to be the prediction of WIMP scattering. We note however that anisotropic distributions, streams, or unvirialised components in the halo phase space may induce changes in the phase of modulation as well as in its functional form~\cite{Green:2003yh,Fornengo:2003fm,Freese:2003na,Savage:2006qr,Green:2010gw,Natarajan:2011gz}.

\section{Model definitions\label{sec:models}}

The main goal of this paper is to evaluate in a quantitative manner the probability of various explanations (i.e., models) for the presence (or absence) of modulation in the CoGeNT time-dependent data, and in particular to estimate the probability of the data being due to a light-mass WIMP signal. To this end, we adopt a phenomenological approach, and, inspired by equation~(\ref{eq:vdep}), parameterise the time-dependent event rate $R(t)$ in the $i$th energy bin as 
\begin{equation}
\label{eq:fit}
R_i(t) = U^i_{\rm m} \left(1 + S^i_{\rm m} \cos[2 \pi (t-t_{\rm max} - 28)/T]\right)\,,
\end{equation}
where $t$ is in units of days,  with $t=0$ corresponding to the first day of data-taking, i.e., December 4, 2009,
$t_{\rm max}$ is the phase of the modulation in terms of days since January 1, and $T$ the modulation period. Two more parameters, $U^i_{\rm m}$ and $S^i_{\rm m}$, denote the mean event rate and the fractional modulation respectively, with the superscript ``$i$'' indicating that these quantities are generally energy-dependent. The binned rates $R_{ij}$ are then simply given by
\begin{equation}
R_{ij} = \frac{1}{\Delta t} \int_{t_j}^{t_j+\Delta_t}   R_i(t) \ \rmd t, 
\end{equation}
where $\Delta t=30$~days.

As discussed in section~\ref{sec:theory}, if the scattering signal is due to dark matter in an isotropic halo, then the modulation phase and period are fixed at $t_{\rm max}=152$~days (i.e., June 2) and
$T = 365$~days respectively, while $U^i_{\rm m}$ and $S^i_{\rm m}$  are determined by the DM mass and cross-section as well as 
the background signal.  In our phenomenological analysis, however, we treat all or subsets of $\{U^i_{\rm m}, S^i_{\rm m},t_{\rm max},T\}$ as free parameters, and 
determine using Bayesian model comparison if the CoGeNT data bear out the dark matter hypothesis or some other alternative scenarios.  We describe the 
various hypotheses we wish to test below; a summary can be found at the end of the section in table~\ref{tab:resume}.

\subsection{Model 0: no modulation}

This is our baseline and also simplest model, representing the case of no modulation in the data, i.e., $S^1_{\rm m}=S^2_{\rm m}=S^3_{\rm m}=0$. In other words, the only relevant and free parameters are  $U^1_{\rm m}$, $U^2_{\rm m}$ and $U_{\rm m}^3$. These are allowed to vary in the ranges indicated in table~\ref{tab:priorsTR}, under uniform priors. The prior ranges have been chosen in such a way that they more than encompass the highest and the lowest measured count rates---including their error bars---in each energy bin. Note that although the prior ranges for these mean rate parameters are somewhat arbitrary, they do not impact on the outcome of our model comparison, because all models considered here have the same $U_{\rm m}^i$ parameters and their prior volume cancels out by virtue of the Savage-Dickey density ratio formula~(\ref{eq:sddr}) (see~\cite{Trotta:2005ar} for details).

\begin{table}[t!]
\caption{Priors for the mean rate $U^i_{\rm m}$ in the three energy bins. All priors are uniform in the indicated ranges.\label{tab:priorsTR}}
\begin{center}
\lineup
\begin{tabular}{ll}
\br
Bin (keVee) &  Prior \\
\mr
$\Delta E_1 =0.5 \to 0.9$   & $20  \to 60$ (counts/30~days) \\
$\Delta E_2= 0.9 \to 3.0$  & $30  \to 100$ (counts/30~days)  \\
$\Delta E_3 = 3.0 \to 4.5$   & $10  \to 60$ (counts/30~days) \\
\br
\end{tabular}
\end{center}
\end{table}

\subsection{Model 1: modulation due to dark matter}

This set of models corresponds to modulation due to a DM signal, where the modulation period is one year, and the phase June 2 is predicted by the scattering of DM in an isotropic halo.  We consider two specific implementations. 

\paragraph{Model 1a: phenomenological DM.} Here we impose a uniform prior in the range $0 \to 0.2$ on the fractional modulation amplitudes $S_{\rm m}^1$ and $S_{\rm m}^2$ in the first two energy bins $\Delta E_1$ and $\Delta E_2$, while for $\Delta E_3$ we assume no modulation, i.e., $S_{\rm m}^3=0$. Such a model encodes the ``na\"{\i}ve'' prediction for the modulated signal due to WIMP scattering in a CoGeNT-like detector~\cite{Drukier:1986tm,Freese:1987wu,Kelso:2010sj}. Compared with model~0, model~1a contains one extra free parameter in each of $\Delta E_1$ and $\Delta E_2$, while there is no change in $\Delta E_3$.

\paragraph{Model 1b: consistent DM signal.} The choice of priors on $S_{\rm m}^i$ reflects our prior belief in the degree of modulation we expect to see in the experimental results under the DM hypothesis. This belief and hence the priors may be updated as we acquire more information from other experiments. In this sense, if our prior belief is that the CoGeNT experiment has indeed detected WIMP scattering events and that the modelling of the background is well understood, we can then use the CoGeNT total event rate to update the prior on the fractional modulation amplitude in each energy bin.

To achieve this purpose, we use Markov Chain Monte Carlo (MCMC) techniques to infer the joint posterior pdf of the DM mass and cross-section---as well as the nuisance and astrophysical parameters---preferred by the CoGeNT total event rate. The DM mass is sampled using a uniform prior in the range $1 \to 20$~GeV, while a uniform prior on $\log (\sigma_n^{\rm SI}/{\rm cm}^2)$ in the range $-41 \to -39$ is assumed for the cross-section. For the DM velocity distribution, we consider an isotropic distribution 
 arising from a Navarro--Frenk--White (NFW) density profile~\cite{Navarro:1996gj}.\footnote{Although it is known that  the Milky Way and spiral galaxies in general are better described by cored profiles~\cite{Salucci:2007tm,Donato:2009ab}, the choice of halo density profile has little impact on the inference results from the current generation of direct DM searches, as shown in~\cite{Arina:2011si}.} Technically, the parametric NFW profile, with its free two parameters $M_{\rm vir}$, the virial mass, and $c_{\rm vir}$, the halo concentration, is ``inverted'' using the Eddington formula under the assumption  of hydrostatic equilibrium. The resulting DM velocity distribution is subject to 
a number of observational and theoretical constraints listed in table~\ref{tab:astro}. For more details see~\cite{Arina:2011si}.
 
\begin{table}[t!]
\caption{Astrophysical constraints on the DM halo profile and the WIMP velocity distribution for model 1b. Except for 
the halo concentration $c_{\rm vir}$ which is subject to a theoretical prior that is uniform in the indicated range,
all other constraints are observational and in the form of Gaussians.\label{tab:astro}}
\begin{center}
\lineup
\begin{tabular}{ll}
\br
Observable & Constraint  \\
\mr
Local standard of rest&  $v_0 = 230 \pm 24.4 \ {\rm km \ s}^{-1}$~\cite{Reid:2009nj,Gillessen:2008qv}\\
Escape velocity &   $v_{\rm esc}= 544   \pm 39 \  {\rm km \ s}^{-1}$~\cite{Smith:2006ym,Dehnen:1997cq} \\
Local DM density & $\rho_{\odot} = 0.4 \pm 0.2 \  {\rm GeV \ cm}^{-3}$~\cite{Weber:2009pt,Salucci:2010qr} \\
Virial mass &  $M_{\rm vir} = 2.7  \pm 0.3 \times 10^{12}\  M_{\odot}$~\cite{Dehnen:2006cm,Sakamoto:2002zr} \\
\mr
Halo concentration & $c_{\rm vir}$ =  $5  \to 20$~\cite{Navarro:2008kc}\\
\br
\end{tabular}
\end{center}
\end{table}

From the joint posterior pdf of $\{m_{\rm DM}, \sigma_n^{\rm SI}, \ldots \}$ one can easily derive posterior pdfs for the fractional modulation amplitudes $S_{\rm m}^i$  
by evaluating for every sample in the Markov Chain the expression
\begin{equation}
\label{eq:permod}
S^i_{\rm m}= \frac{R^{\rm DM+bckg}_i({\rm June\  2})- R^{\rm DM+bckg}_i({\rm December  \ 2})}{R^{\rm DM+bckg}_i({\rm June \ 2}) + R^{\rm DM+bckg}_i({\rm December \ 2})}\,,
\end{equation}
where $R_i^{\rm DM+bckg}(t)$ is the combined DM and background event rate expected in the $i$th energy bin, and June~2 (December~2) is the day when the WIMP scattering rate is expected to reach a maximum (minimum). Importantly, the background event rate, as given in equation~(\ref{eq:bckg}), is treated as a constant component in time. The resulting chains in $S_{\rm m}^i$ are  binned to construct the posterior pdfs, which are in turn the priors we seek for the present modulation analysis.  Note that because we are using {\it only} the total rate to derive a prior for $S_{\rm m}^i$ and no time information at this stage, we are not fitting the same data twice in the  procedure. 

\begin{figure}[t!]
\includegraphics[width=1.\columnwidth]{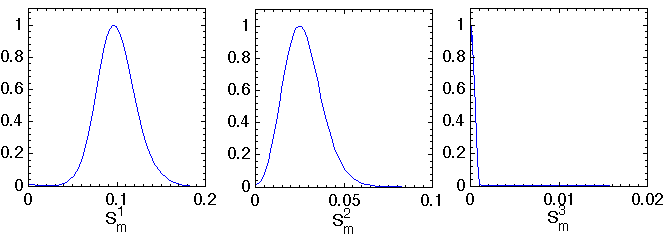}
\caption{Priors for the fractional modulation amplitudes $S^i_{\rm m}$ in model 1b, derived from the CoGeNT total rate.\label{fig:prior1b}}
\end{figure}
\begin{figure}[h]
\begin{minipage}{0.49\textwidth}
\centering
\includegraphics[width=1.\columnwidth]{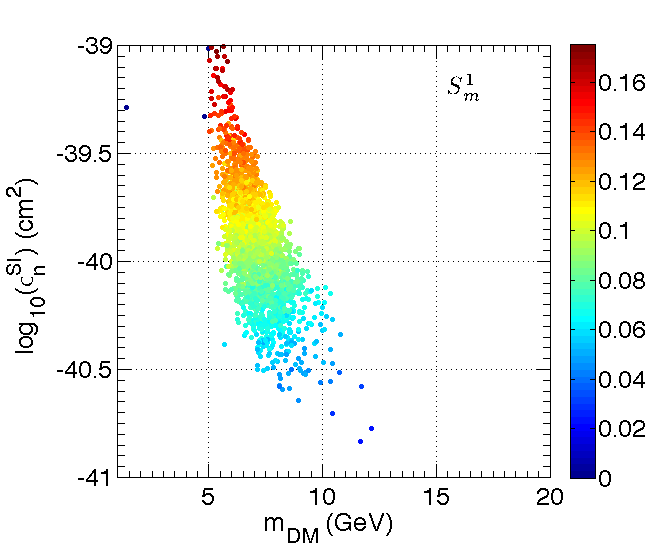}
\end{minipage}
\begin{minipage}{0.49\textwidth}
\centering
\includegraphics[width=1.\columnwidth]{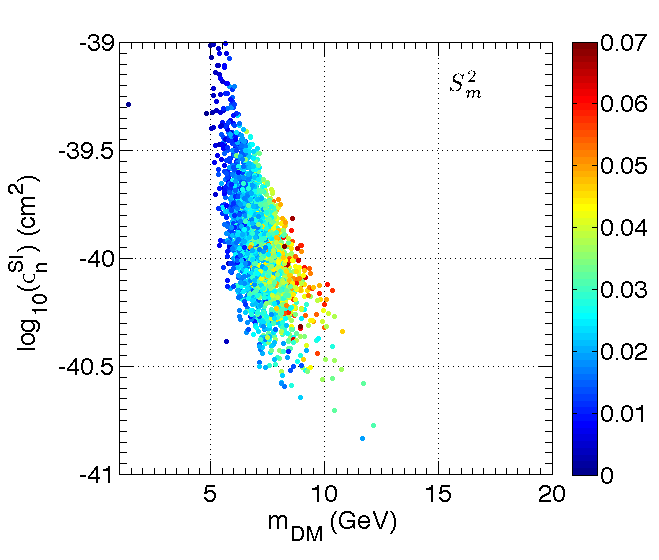}
\end{minipage}
\\
\begin{minipage}{0.49\textwidth}
\centering
\includegraphics[width=1.\columnwidth]{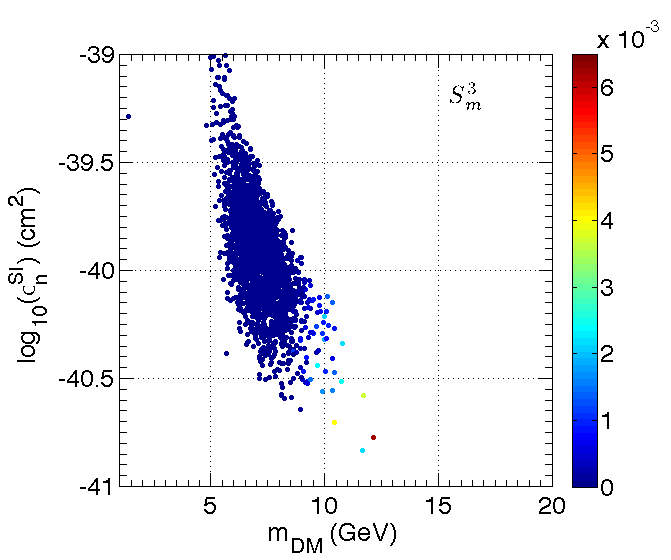}
\end{minipage}
\begin{minipage}{0.49\textwidth}
\caption{Samples from the posterior pdf in the $\{m_{\rm DM}, \sigma_n^{\rm SI},S_{\rm m}^i\}$ plane, where the $S_{\rm m}^i$ value is indicated by the colour coding.
The top panels show $S_{\rm m}^1$ (left) and $S_{\rm m}^2$ (right), while the bottom panel shows $S_{\rm m}^3$.\label{fig:prior1bcorrelations}}
\end{minipage}
\end{figure}

In the case we consider, inclusion of astrophysical nuisance parameters in deriving effective priors for $S_{\rm m}^i$ leads to the width of such priors being larger than it would otherwise have been, had the astrophysical quantities been kept fixed. This in turn means that the evidence value for model 1b will be reduced by virtue of the Occam's razor effect with respect to the case in which the astrophysical parameters are held fixed. In this sense, our evidence values for model 1b are to be considered conservative.

Figure~\ref{fig:prior1b} shows the priors on $S^i_{\rm m}$ thus constructed for the three energy bins, while figure~\ref{fig:prior1bcorrelations}
shows the correlations between $S_{\rm m}^i$, the DM mass, and the cross-section. In the case of the first bin, the fractional modulation amplitude $S_{\rm m}^1$ increases with  the cross-section, while its dependence on the DM mass is weak. In contrast, $S_{\rm m}^2$ in the second bin depends strongly on the DM mass, with higher masses corresponding to larger modulation amplitudes. These observations indicate that $S_{\rm m}^1$ and $S_{\rm m}^2$ are minimally correlated with one another, and thereby justify our treatment of them as independent parameters in the model comparison analysis. Finally, $S_{\rm m}^3$ in the third bin is close to zero, since for WIMPs with masses smaller than $\sim 10$~GeV, the exponentially decaying DM scattering signal becomes vanishingly small. 

\subsection{Model 2: modulation due to some other physics}

If the time modulation is not due to WIMP scattering off nuclei but to some unknown physics, then the phase $t_{\rm max}$ need not correspond to June~2, or the period $T$ exactly one year. In this case, then, $t_{\rm max}$ and $T$ should also be treated as free parameters alongside $S_{\rm m}^i$ and $U_{\rm m}^i$. We consider two distinct scenarios.
 
\paragraph{Model 2a: non-DM, annual modulation.} This model has an annual modulation, but with its modulation amplitudes and phase treated as free parameters. For the latter, we vary $t_{\rm max}$ within $0 \to 365$~days under  a uniform prior. Note that unlike the parameters $S_{\rm m}^i$ and $U_{\rm m}^i$ which can change from energy bin to energy bin, there is only one $t_{\rm max}$ parameter in the analysis. In other words, in a combined analysis of all three energy bins, the time modulations in the expected signals are described by the same phase.
 
\begin{table}[t!]
\caption{Summary of all models studied here. Unless otherwise stated, all priors are uniform in the indicated ranges. The quantity $\nu$ counts the number of extra free parameters in a model with respect to model~0, with the first number referring to the one-bin analysis and the second to the all-bins analysis. \label{tab:resume}}
\begin{center}
{\small
\begin{tabular}{|ll|llll|}
\hline
\hline
Model  & Description &  Fractional  &  Phase $t_{\rm max}$ & Period $T$ & Extra\\
 & &modulation $S_{\rm m}^i$ & (days) & (days) & params\\
\hline
\hline
 0 & No modulation  & 0  & --- & --- & $\nu=0,0$\\
 \hline
1a  & Pheno-DM & $S_{\rm m}^{1,2} = \ 0 \to 0.2$ & 152 &365 &$\nu =1,2$\\ 
       &                  & $S_{\rm m}^3=0$  & & &\\
1b & Consistent DM & Gaussian, clipped at 0&152 & 365 & $\nu =1,3$\\
& & ($S_{\rm m}^i \geq 0$) &&& \\
    & & $S_{\rm m}^1 =0.098\pm 0.021$ & & &\\
      &  & $S_{\rm m}^2 =0.026\pm0.011$ & & &\\
        & & $S_{\rm m}^3 =(0.37\pm 36)\times 10^{-4}$ & & &\\
    \hline
  2a  & Non-DM, annual & $0 \to 1$  & $0 \to 365$ & 365 & $\nu=2,4$\\
  2b& Non-DM, free period & $0 \to 1$  & $0 \to 365$  &  $1 \to 365$ & $\nu=3,5$\\
\hline
\hline
\end{tabular}
}
\end{center}
\end{table}

\paragraph{Model 2b: non-DM, free period.} This is similar to model~2a, but in addition, the period $T$ is allowed to vary in the range $1 \to 365$~days under a uniform prior. As with the parameter $t_{\rm max}$, we use only one $T$ parameter in the analysis.

\bigskip
 
\noindent All models under consideration are summarised in table~\ref{tab:resume}.

\section{Results and discussions\label{sec:res}}

\begin{figure}[t!]
\centering
\includegraphics[width=1.\columnwidth, trim=0mm 10mm 25mm 20mm, clip]{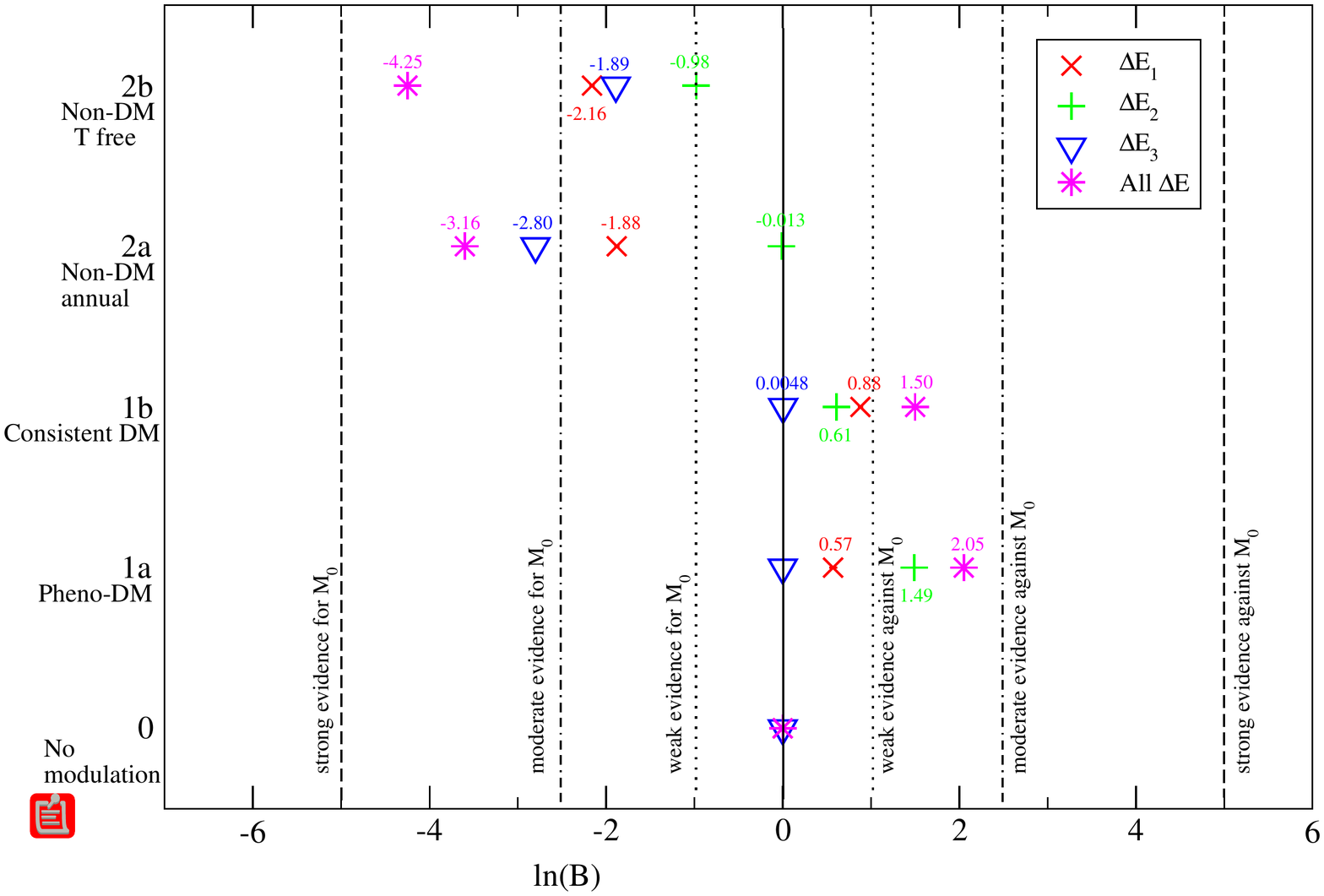}
\caption{Bayes factors for the various modulation scenarios analysed in this work. The models are specified on the vertical axis, 
while the different symbols refer to the energy bin(s) for which the Bayes factors have been computed, as labelled in the plot. The actual value of $\ln B$ in each case is indicated by the number above the data point. The Bayes factors have uncertainties of $\sim 0.02$ for the individual bins and $\sim 0.04$ for the combined analysis. Following Jeffrey's scale in table~\ref{tab:jef}, the vertical lines demarcate the different empirical gradings of the strength of the evidence.
\label{fig:lnB}}
\end{figure}

The main results of our analysis are the Bayes factors, relative to model~0, for models~1~and~2. These are displayed in figure~\ref{fig:lnB}, and 
the corresponding odds for each model against model~0 are listed in table~\ref{tab:odds}. For reference, we show also the difference in twice the best-fit log-likelihood value, $\Delta \chi_{\rm eff}^2$, and, where analytically calculable, the p-values of the null in table~\ref{tab:bmax}. We elaborate on the salient features of our results below.

\begin{figure}[t!]
\begin{minipage}[t]{0.32\textwidth}
\centering
\includegraphics[width=1.1\columnwidth,trim=0mm 10mm 0mm 15mm, clip]{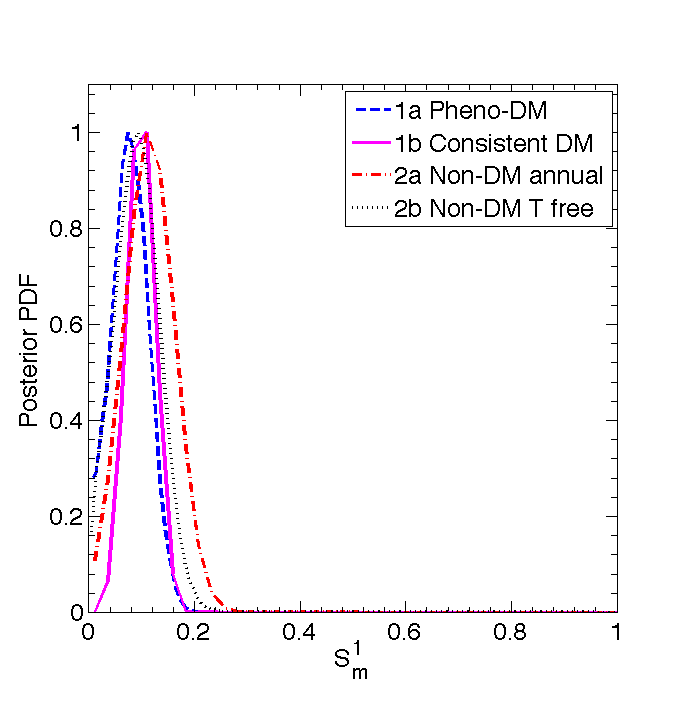}
\end{minipage}
\begin{minipage}[t]{0.32\textwidth}
\centering
\includegraphics[width=1.1\columnwidth,trim=0mm 10mm 0mm 15mm, clip]{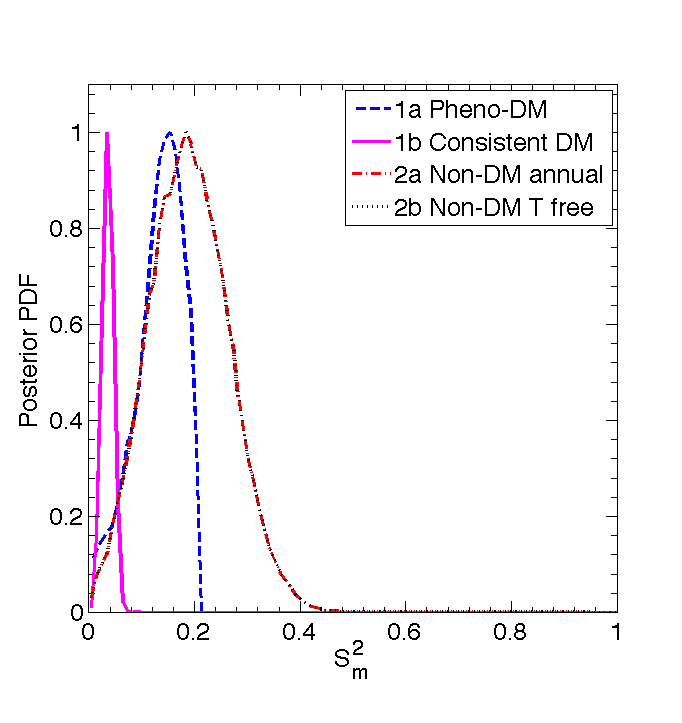}
\end{minipage}
\begin{minipage}[t]{0.32\textwidth}
\centering
\includegraphics[width=1.1\columnwidth,trim=0mm 10mm 0mm 15mm, clip]{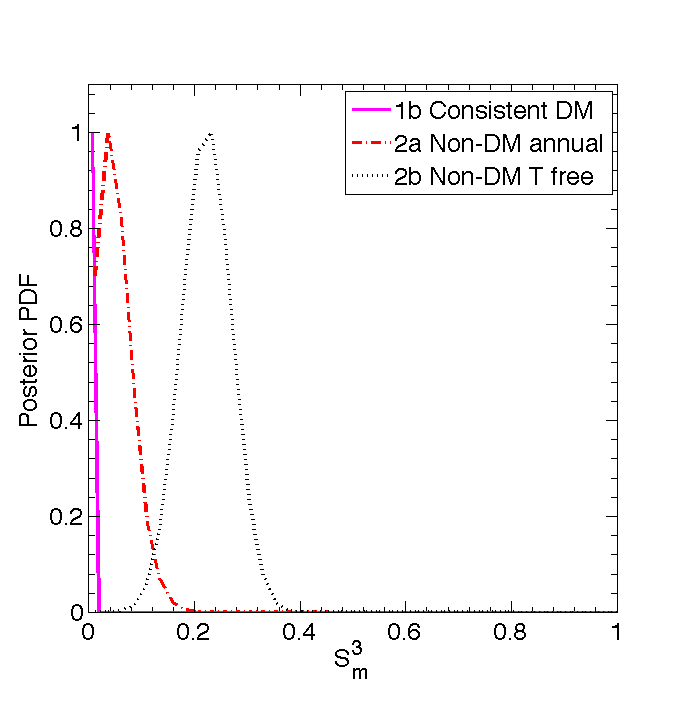}
\end{minipage}
\\
\begin{minipage}[t]{0.32\textwidth}
\centering
\includegraphics[width=1.1\columnwidth,trim=0mm 10mm 0mm 15mm, clip]{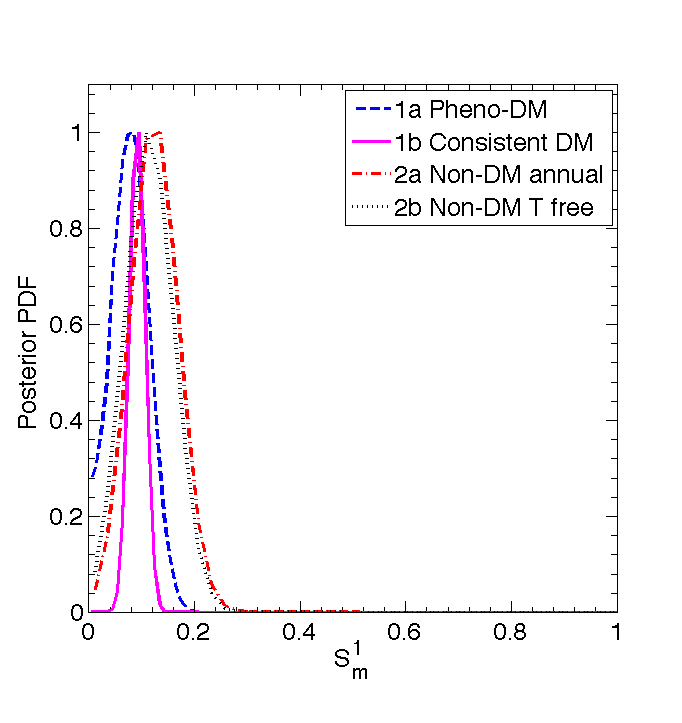}
\end{minipage}
\begin{minipage}[t]{0.32\textwidth}
\centering
\includegraphics[width=1.1\columnwidth,trim=0mm 10mm 0mm 15mm, clip]{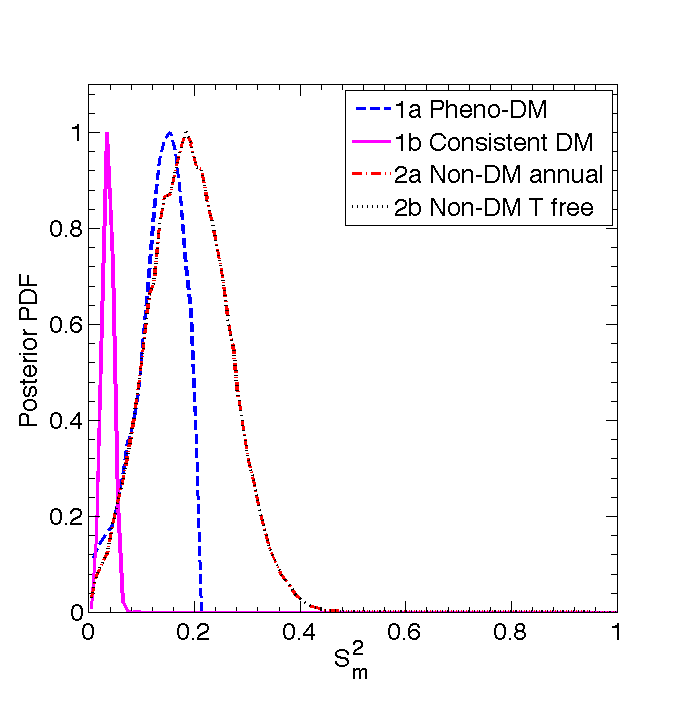}
\end{minipage}
\begin{minipage}[t]{0.32\textwidth}
\centering
\includegraphics[width=1.1\columnwidth,trim=0mm 10mm 0mm 15mm, clip]{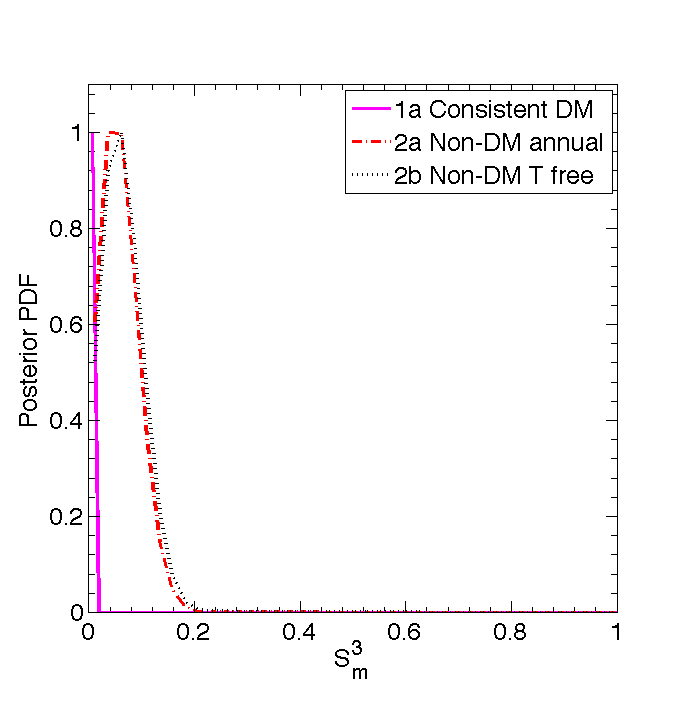}
\end{minipage}
\caption{1D marginal posterior pdfs (normalised to the peak) for, from left to right, the modulation amplitudes $S_{\rm m}^1$, $S_{\rm m}^2$ and $S_{\rm m}^3$ in the 3 CoGeNT energy bins, for all models considered in this work. Results from the single bin analyses are displayed in the top panels, while the bottom panels show results from the combined analysis, which includes all 3 energy bins in the likelihood.
\label{fig:Smpdf}}
\end{figure}

\subsection{Bin-by-bin description}

\paragraph{Energy bin 1} 

As shown in figure~\ref{fig:lnB}, the evidence for modulation in $\Delta E_1$ is generally inconclusive for DM models (models~1), and weakly against ``other physics'' models (models~2), when compared with the no modulation model.
This yields a moderate to strong evidence for DM models as compared to the ``other physics'' models.
Of the DM models, model~1b (DM signal consistent with CoGeNT total rate) is marginally favoured over model~1a (phenomenological DM model), despite the fact that purely from the quality of fit  (see table~\ref{tab:bmax}), model~1b performs marginally worse than~1a. This is an example of Lindley's paradox (i.e., Bayesian model selection returning a different result from classical hypothesis testing, see~\cite{Trotta:2005ar} and references therein):  model~1b is rewarded with a larger Bayes factor because of its narrow prior,  $S_{\rm m}^1=0.098 \pm 0.021$, which makes it highly predictive compared with model~1a (prior  $S_{\rm m}^1=0 \to 0.20$).

\begin{figure}[t!]
\begin{minipage}[t]{0.32\textwidth}
\centering
\includegraphics[width=1.1\columnwidth, trim=5mm 10mm 0mm 5mm, clip]{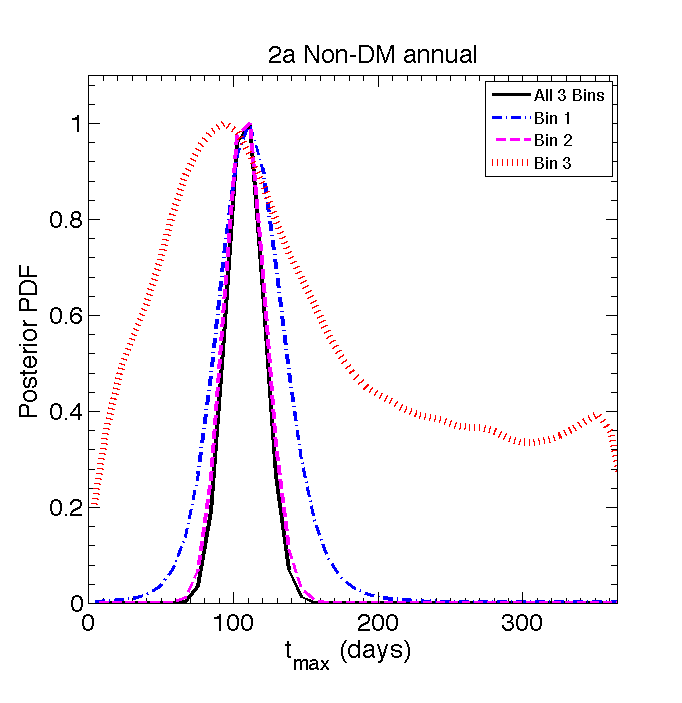}
\end{minipage}
\begin{minipage}[t]{0.32\textwidth}
\centering
\includegraphics[width=1.1\columnwidth, trim=5mm 10mm 0mm 5mm, clip]{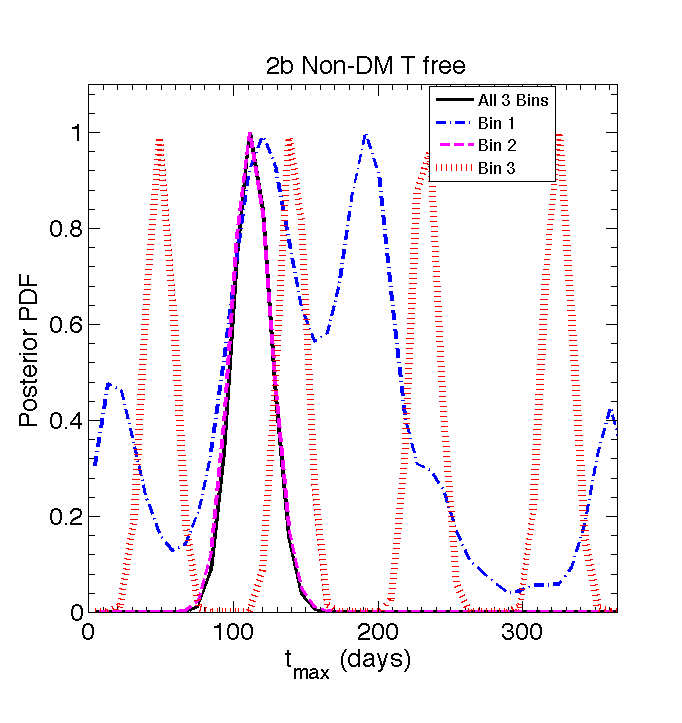}
\end{minipage}
\begin{minipage}[t]{0.32\textwidth}
\centering
\includegraphics[width=1.1\columnwidth, trim=5mm 10mm 0mm 5mm, clip]{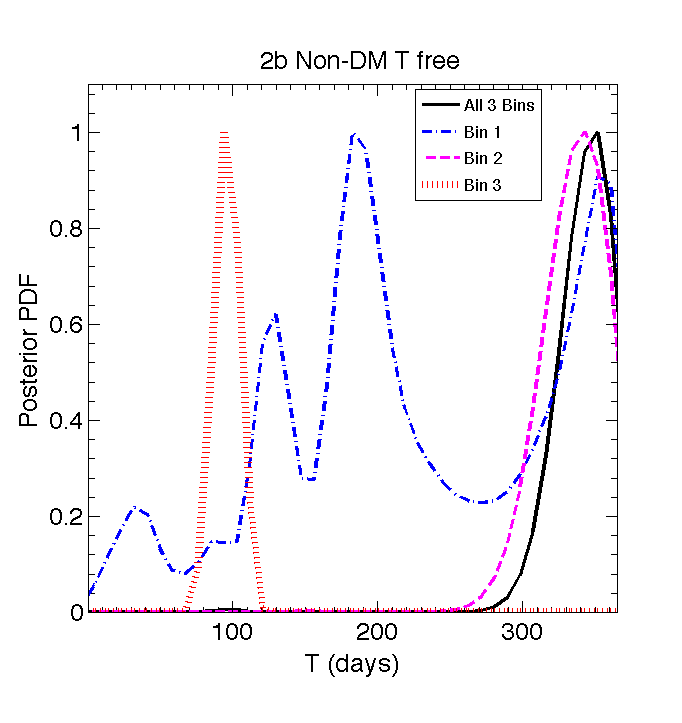}
\end{minipage}
\caption{1D marginal posterior pdfs for the phase $t_{\rm max}$ in models 2a and 2b (left and middle panel respectively), and 
and for the period $T$ in model 2b (right panel) from the single bin and the combined analyses, as labelled in the plots. 
\label{fig:Tpdf}}
\end{figure}

\begin{figure}[t!]
\begin{minipage}[t]{0.49\textwidth}
\centering
\includegraphics[width=1.1\columnwidth,trim=0mm 10mm 0mm 15mm, clip]{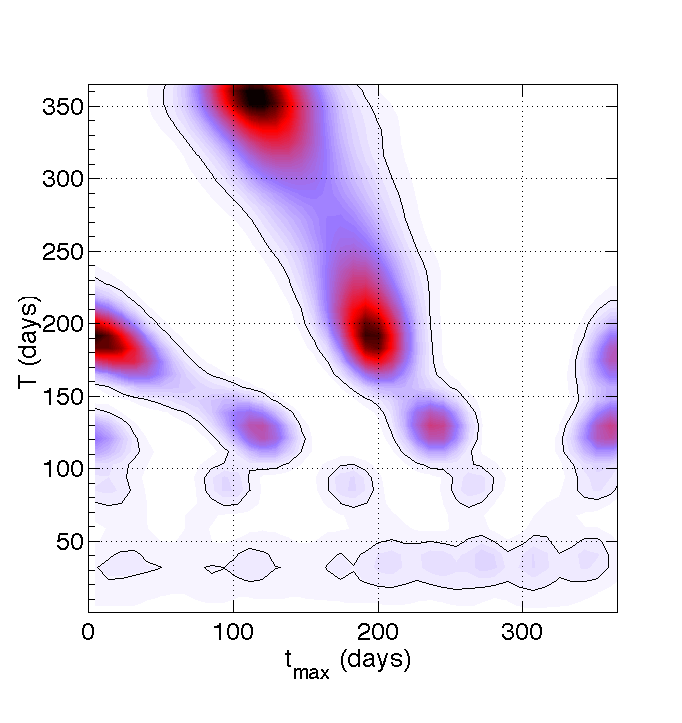}
\end{minipage}
\begin{minipage}[t]{0.49\textwidth}
\centering
\includegraphics[width=1.1\columnwidth,trim=0mm 10mm 0mm 15mm, clip]{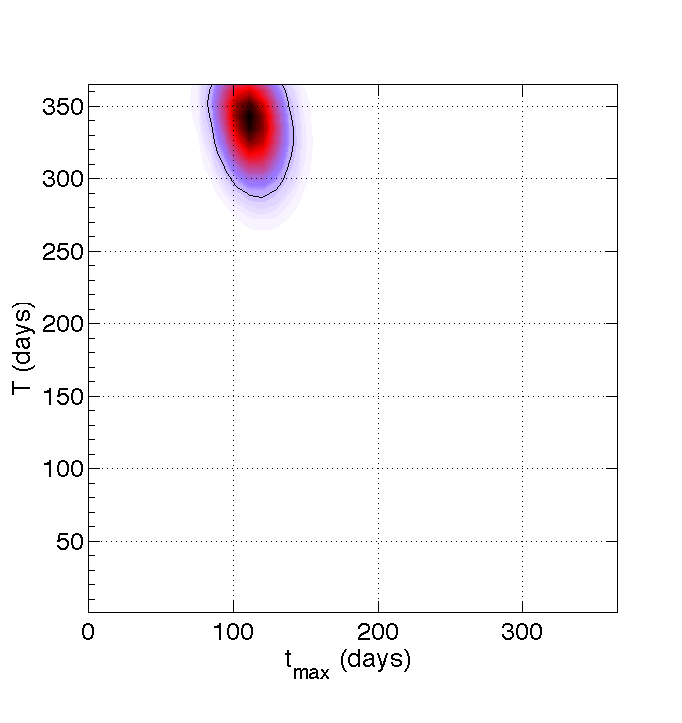}
\end{minipage}
\\
\begin{minipage}[t]{0.49\textwidth}
\centering
\includegraphics[width=1.1\columnwidth,trim=0mm 10mm 0mm 15mm, clip]{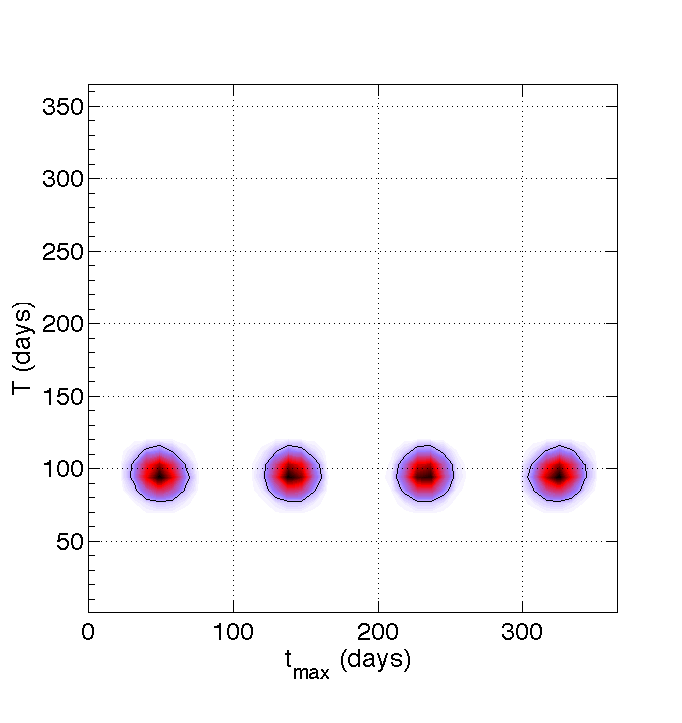}
\end{minipage}
\begin{minipage}[t]{0.49\textwidth}
\centering
\includegraphics[width=1.1\columnwidth,trim=0mm 10mm 0mm 15mm, clip]{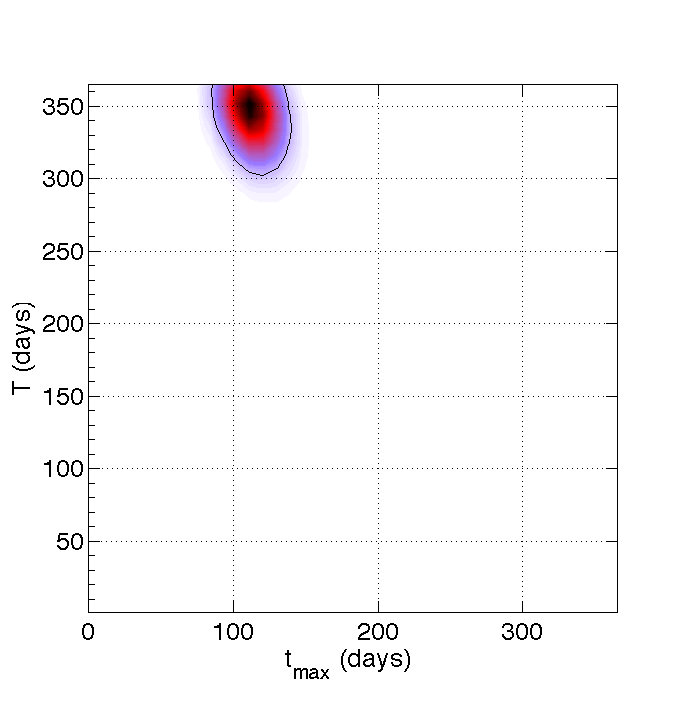}
\end{minipage}
\caption{2D marginal posterior pdfs in the $\{t_{\rm max},T \}$-plane for model~2b from the single-bin analyses---bin~1 (top left), bin~2 (top right), and
bin~3 (bottom left)---and for the combined fit (bottom right).
The black solid lines are 90\%-contours.\label{fig:2DTtmax}}
\end{figure}

The posterior marginal distribution for the fractional modulation amplitude in each scenario is shown in the top left panel of figure~\ref{fig:Smpdf}. Observe that all DM scenarios select comparable values for the modulation amplitude, e.g., model~1a prefers $S_{\rm m}^1 = 0.08 \pm 0.03$ and model~1b $S_{\rm m}^1 = 0.09 \pm 0.01$, because in both cases the phase has been fixed to June~2 ($t_{\rm max} \sim 152$~days). However, if the phase is let free to vary, model 2a selects a slightly larger value for the modulation amplitude, $S_{\rm m}^1 = 0.11 \pm 0.04$, and a preferred phase $t_{\rm max} = 106 \pm 29$~days, as shown in figure~\ref{fig:Tpdf}. Model~2b is heavily multimodal both in $t_{\rm max}$ and the period $T$ (see also figure~\ref{fig:2DTtmax}, top left panel).  Because of this multimodality, we do not summarise the inference for $t_{\rm max}$ and $T$ in model~2b
in terms of the 1D posterior mean and standard deviation.

\paragraph{Energy bin 2}  Compared with model~0, the support for modulation in energy bin~2 is at best weak.  With the exception of the moderate evidence for model~1a against model~2b,
 the comparisons between the modulation models themselves are likewise weak to inconclusive.
The preferred fractional modulation amplitude is $S_{\rm m}^2= 0.20 \pm 0.05$ for both models~2a and~2b,
while for model~1a the posterior mode is close to the prior boundary of $0.2$ (see figure~\ref{fig:Smpdf}). By comparing the posterior for $S_{\rm m}^2$ for models 2a and 2b with that for models 1a and 1b, it is clear that the data in energy bin 2 prefer a larger modulation amplitude that can be provided by either DM models. 
The posterior pdfs for both models~2a and~2b are unimodal in $t_{\rm max}$ and, for model 2b, $T$, giving $t_{\rm max}=104 \pm 11$~days for model~2a, and 
$t_{\rm max} = 108 \pm 15$~days and $T = 331\pm 23$~days for model~2b, as shown in figures~\ref{fig:Tpdf} and~\ref{fig:2DTtmax}.

\paragraph{Energy bin 3}

For WIMP masses smaller than 10~GeV, the generic prediction for CoGeNT-like detectors is that there should be 
little to no annual modulation in the $3.0 \to 4.5$~keVee energy range.   Indeed, we find in this energy bin that the no modulation scenario is au pair with the DM models, and weakly to moderately preferred over the modulation due to ``other physics'' models.
It is interesting to note that the evidence against model~2b relative to model~0 is only weak.  This is because with its freely varying phase $t_{\rm max}$ and period $T$, model~2b turns out to provide a better fit to the time-structure of the CoGeNT data (left panel of figure~\ref{fig:CoGSmt}) than a constant rate. 
As can be seen in figures~\ref{fig:Tpdf} and~\ref{fig:2DTtmax}, the posterior pdf is unimodal in the period $T$, with a preference for $T= 91 \pm 4$~days, but is strongly multimodal in the direction of the phase~$t_{\rm max}$.

Thus, we conclude that in the energy range $3.0 \to 4.5$~keVee, the CoGeNT data do not support the presence of modulation.
This statement is reinforced by the behaviour of the 1D marginal posterior for $S_{\rm m}^3$ in the top right panel of figure~\ref{fig:Smpdf}, which shows a modulation amplitude consistent with zero in models~1b and~2a.

\subsection{All bins analysis: dark matter or other physics?}

When the data from all three energy bins are analysed simultaneously, we find that only the DM models~1a and~1b receive weak support from the data over no modulation.
In contrast, models~2a and~2b are moderately disfavoured with respect to the no modulation scenario. DM models are thus strongly favoured over ``other physics" models, with odds in favour of the DM models ranging from $185:1$ to $560:1$, see table~\ref{tab:odds}. This is a consequence of the predictiveness of models~1a and~1b; that is, Occam's razor is at work, penalising the ``other physics" models for their excessive  free parameters unsupported by the data. Table~\ref{tab:odds} summarises the odds for each model against the no modulation model~0. Odds between any pair of models can be obtain simply by multiplication of the odds given there.

The inferred values for $S_{\rm m}^i$  are closely in line with the single bin analysis (see figure~\ref{fig:Smpdf}), with the exception of bin 3 in model~2b, which now prefers $S_{\rm m}^3 = 0.03 \pm 0.04$ in the combined fit, instead of $S_{\rm m}^3 = 0.22 \pm 0.04$ when the bin is analysed on its own.
 This result comes about because the phase $t_{\rm max}$ and the period $T$ are constrained to be the same in all three bins in the combined fit. The fit, in turn, is driven mainly by the data in bins~1 and~2. Indeed, the common phase preferred by the combined data is now $t_{\rm max}= 104\pm 10$~days for model~2a and $107 \pm 12$~days for model~2b, while the preferred period is $T= 344 \pm 22$~days in model~2b (see also figure~\ref{fig:Tpdf}). Figure~\ref{fig:2DTtmax} also shows that the all-bin fit for $t_{\rm max}$ and $T$ is dominated by bin 2, which singles out a value $t_{\rm max} \sim 100$ and $T\sim 345$ from the vast degenerate region obtained from bin 1 alone. On the other hand, the multi-modal posterior preferring $T \sim 100$ days from bin 3 is completely cut away in the combined fit. This shows that bin 3 on its own prefers value for phase and period that are incompatible with bins 1 and 2. 

It is important, like in any good Bayesian analysis, to assess the robustness of the results with respect to reasonable changes in our prior choices. An interesting question to ask in this regard is, have the Bayes factors for models~2a and~2b been artificially suppressed because of our choice of priors? To answer this question, we appeal to the Savage-Dickey density ratio defined in equation~(\ref{eq:sddr}).  Consider for concreteness model~2a. If we were to reduce the prior ranges for all three fractional modulation amplitudes to $S_{\rm m}^i = 0 \to 0.5$ from the default choice of $S_{\rm m}^i=0 \to 1$, then it follows from the SDDR formula that the Bayes factor 
 $\ln B$ in favour of model 2a would increase by approximately $\ln 2^3 \simeq 2.1$~units, bringing it to $\sim -1.06$ relative to model~0. In such a case, the model comparison between model~2a and the consistent DM model~1b would produce a weak evidence in favour of the latter scenario, while a comparison between models 2a and 1a would still favour moderately the latter.  Indeed, even with the prior ranges reduced further to  $S_{\rm m}^i = 0 \to 0.2$, model~2a would only barely overtake model~1b, and would still lag marginally behind model~1a as the most preferred model by the data.  Therefore, we conclude that the general statement that DM-like models are preferred over ``other physics'' models is robust from a Bayesian point of view.

\begin{table}[t!]
\caption{Odds between modulation models $\mathcal{M}_i$ versus the no modulation model $\mathcal{M}_0$ for the CoGeNT modulation data. \label{tab:odds}}
\begin{center}
\lineup
\begin{tabular}{|c|llll|}
\hline
\hline
& \multicolumn{4}{c|}{$\mathcal{M}_i: \mathcal{M}_0$} \\
 Model ${\mathcal M}_i$&  Bin 1 &  Bin 2 &  Bin 3 &  All 3 bins\\
 \hline
 \hline
 1a & $2:1$& $4: 1$	 & $1:1$ &$8:1$ \\
1b & $2 :1$ & $2:1$ & $1:1 $ & $5:1$ \\
 \hline
 2a &$1:7$	& $1 :1$ & 	$1:16$	&$1:37$ \\
2b &$1:9$	& $1:3$& $1:6$ & $1:70 $ \\
 \hline
 \hline
 \end{tabular}
\end{center}
\end{table}

\subsection{Comparison with classical p-values}\label{sec:pval}

In frequentist statistics, classical hypothesis testing seeks to rule out the null hypothesis $H_0$ by quantifying the probability of observing data as extreme or more extreme than what has been obtained. To this end, a test statistic $t$ can be defined, in such a way that extreme values of $t$ are increasingly improbable under $H_0$. 
For concreteness, we  assume that, under the null, larger values of $t$ have monotonically decreasing probabilities. Then the tail probability, or the  p-value $\wp$, can be computed via
\begin{equation}
 \wp \equiv \int_{t_{\rm obs}}^\infty p(t|H_0),
\end{equation}
so that small values of $\wp$ denote that the observed data are very improbable under the null. We stress once more that p-values are {\em not} probabilities for hypotheses---they are probabilities of obtaining more extreme data than observed {\em assuming the null hypothesis is correct}. In order to obtain the probability for a hypothesis (which is arguably the scientific question we are interested in), one needs to take a Bayesian approach, as we do in this work. Also, we caution that the mapping of the test statistic $t$ onto a p-value requires in general a Monte Carlo simulation; analytic solutions exist only in special cases, and apply {\it only} if certain regularity conditions hold, as we discuss below.

A popular choice for the test statistic $t$ in the context of nested models is the effective chi-square (or more precisely, the profile likelihood ratio),%
\footnote{Strictly speaking, the quantity we compute for model~1b is the posterior ratio, since the prior pdfs in that case are not uniform.}
\begin{equation} \label{eq:chieff} 
\Delta \chi^2_{\rm eff} \equiv -2 \ln \left[ \frac{{\mathcal L}(\vartheta^\star,\hat{\psi})}{{\mathcal L}(\hat{\hat{\vartheta}},\hat{\hat{\psi}})} \right]\,, 
\end{equation}
where ${\mathcal L}(\vartheta^\star,\hat{\psi})$ is the conditional maximum likelihood fixing $\vartheta = \vartheta^\star$, and ${\mathcal L}(\hat{\hat{\vartheta}},\hat{\hat{\psi}})$ is the unconditional maximum likelihood in the whole parameter space of the more complex model. 
 Let us identify the simpler model (with $\vartheta = \vartheta^\star$) as the null hypothesis that we seek to rule out (e.g., no modulation in the CoGeNT data).  If the likelihood in the $N$ additional parameters of the more complex model is Gaussian and unbounded, then Wilks'
theorem~\cite{Wilks:1938} applies, meaning that the test statistic $\Delta \chi^2_{\rm eff}$ is asymptotically distributed as a $\chi^2$ with $N$ degrees of freedom. We note however that one of the conditions that validate the application of  Wilks' theorem is that the likelihood must be unbounded, i.e., the additional parameters of the more complex model cannot sit on the boundary of its parameter space. This is precisely the situation we encounter when treating the modulation amplitudes $S_{\rm m}^i$, where $S_{\rm m}^i = 0$
under the null hypothesis. 
Hence, Wilks' theorem cannot be applied to obtain the p-value from the $\Delta \chi^2_{\rm eff}$, a procedure that is often followed outside its realm of validity (see~\cite{Protassov:2002sz} for a discussion aimed at astronomers).

\begin{table}[t!]
\caption{$\Delta \chi^2_{\rm eff}$ values, as defined in equation~(\ref{eq:chieff}), for the various modulation scenarios relative to model~0. Where analytically calculable, we quote also the corresponding classical p-values of the null hypothesis (model~0), obtained via Chernoff's theorem, Eq.~\ref{eq:chernoff}. We also give the number of extra free parameters in the alternative  hypothesis relative to the null.\label{tab:bmax}}
\begin{center}
\lineup
\begin{tabular}{|c|llll|}
\hline
\hline
& \multicolumn{4}{c|}{$\Delta\chi^2_{\rm eff}$  relative to model~0} \\
 Model &  Bin 1 &  Bin 2 &  Bin 3&  All 3 bins\\
 \hline
 \hline
 1a & $2.04$ & $4.23$	 & -- &	$6.26$ \\
  &$\wp = 0.08 $ &$\wp = 0.02$  & -- &$\wp = 0.02$ \\
  & ($\nu=1$) & ($\nu=1$) & & ($\nu=2$) \\
   1b & $1.94$ & $1.88$	 & $0.020$ &	$3.84$ \\
 & $\wp = 0.08$ & $\wp = 0.09$  &$\wp = 0.4$ &$\wp = 0.1$\\
  & ($\nu=1$) & ($\nu=1$) & ($\nu=1$) & ($\nu=3$) \\
 \hline
 2a & $3.61$	& $8.36$ & 	$0.025$	&$10.63$ \\
2b & $3.70$ 	& $8.87$ & 	$10.88$	& $10.86$ \\
 \hline
 \hline
 \end{tabular}
\end{center}
\end{table}

For the case where (i)~the $N$ additional parameters are bounded, with the null hypothesis sitting on the boundary, (ii)~the likelihood is Gaussian, and (iii)~all parameters are identifiable under the null hypothesis, we can use instead Chernoff's theorem~\cite{Chernoff:1954,Shapiro:1988} to evaluate the p-value of the null. Chernoff's theorem says that the distribution of the test statistic $\Delta \chi^2_{\rm eff}$ under the null is asymptotically a weighted sum of random variables $\chi^2_i$ following chi-squared distributions with $i$ degrees of freedom:
\begin{equation} \label{eq:chernoff}
	\wp = \sum_{i=0}^N 2^{-N} {N \choose i} p(\chi^2_i > \Delta
	\chi^2_{\rm eff}). 
\end{equation}
Equation~(\ref{eq:chernoff}) can be used to compute the p-value of the null hypothesis of no modulation when the more complex hypothesis is identified with models~1a or~1b, i.e., the DM hypotheses. However, it cannot be applied to the ``other physics'' models~2a and~2b, for these models contain parameters (the phase and the period) that are undefined and unidentifiable under the null, i.e., when $S_{\rm m}^i = 0$ in models 2a and 2b, the parameters $t_{\rm max}$ and $T$ are meaningless. 

Applying equation~(\ref{eq:chernoff}), we obtain a p-value of $0.02$ under the null for the three energy bins combined, when the alternative hypothesis is model 1a. This formally corresponds to a $2.3\sigma$ detection (using a Gaussian distribution to convert p-values into the number of sigmas). If instead we take model 1b as the alternative, the p-value is $0.1$, equivalent to a $1.6\sigma$ detection. Monte Carlo simulations would be required to determine the distribution of the test statistic when the alternative model is either 2a or 2b.

\subsection{Impact of an anisotropic velocity distribution}

Although DM models appear to be preferred over modulation models due to other physical processes, there remains some conflict between the  modulation amplitude 
preferred by the CoGeNT time-dependent data in energy bin~2 and the amplitude predicted by the experiment's total event rate. Therefore, we wish to investigate whether there is a WIMP scenario that could reconcile these findings. Modifications to the particle physics interaction have been widely discussed in the context of, e.g., inelastic or isospin violating DM models~\cite{Farina:2011pw,Schwetz:2011xm,Fox:2011px,Hooper:2011hd,Gao:2011ka,Rajaraman:2011wf,An:2011ck,McCabe:2011sr}. Here, we consider the possibility of an anisotropic DM velocity distribution in the Galactic halo. Anisotropic velocity distributions have been investigated in~\cite{Belli:2011kw} as a means to reconcile the CoGeNT and the DAMA data, and more recently in~\cite{Frandsen:2011gi} to reconcile both the CoGeNT total rate with the CRESST excess and the CoGeNT modulation with its total rate.

Consider a class of ellipsoidal (or equivalently, triaxial) halo models whose gravitational potential is given by 
$\Phi(x,y,z)=(v_c/2) \ln(x^2+y^2/p^{2}+z^2/q^{2})$, where $x,y,z$ are the (Cartesian) coordinates, and $q^2 \leq p^2 \leq 1$ parameterise the degree of triaxiality~\cite{Evans:2000gr}. Assuming the Earth sits on either the long axis ($x$-axis) or the intermediate ($y$-axis), the corresponding DM velocity distribution is in the form of a triaxial Gaussian, i.e., 
\begin{equation}
\label{eq:triax}
f(\vec{v'}(t)) = \frac{1}{(2 \pi)^{3/2} \sigma_R \sigma_\phi \sigma_z} \exp\left[-\frac{{v'}^2_R}{2 \sigma_R^2}
-\frac{(v'_\phi+v_\oplus)^2}{2 \sigma_\phi^2} - \frac{{v'}^2_z}{2 \sigma_z^2}\right], 
\end{equation}
in the Earth's rest frame, where $v_\oplus$ is the Earth's speed in the Galactic frame given in equation~(\ref{eq:voplus}), and
$\sigma_X$, with $X=R,\phi,z$, is the velocity dispersion in the $X$-direction in a cylindrical coordinate system. The exact expressions for $\sigma_{R,\phi,z}$ can be found in equations~(4.25) and~(4.26) of~\cite{Evans:2000gr}: specifically, they depend on $v_0$ and on the anisotropy parameter $\gamma$. In the spherical limit, equation~(\ref{eq:triax}) reduces to the well-known Gaussian velocity distribution of singular isothermal halos.%
\footnote{It is a reasonable approximation to consider 
an isothermal halo and its ellipsoidal extensions, 
because direct detection is sensitive only to the local properties of the velocity distribution and
cannot presently distinguish between different DM density profiles~\cite{Arina:2011si,Arina:2011xu}.}

One possibility is a radial anisotropy, $\sigma_R > \sigma_z = \sigma_\phi$, or equivalently, $\gamma <0$. However, this kind of anisotropy does not give the desired effect of enhancing the fractional modulation amplitude, because the WIMP phase space is partially depleted in the 
direction of the Sun's movement around the Galactic centre, thereby reducing the fractional difference between the maximum and the minimum event rates relative to the mean~\cite{Belli:2002yt}. In contrast, a tangential anisotropy $\gamma > 0$ may potentially enhance the fractional modulation amplitude~\cite{Evans:2000gr,Belli:2002yt,Fornengo:2003fm}.

We perform a MCMC for {\it one specific} realisation of a tangentially anisotropic triaxial halo (anisotropy parameters $p=0.9$, $q=0.8$ and $\gamma = 16$, and astrophysical parameters fixed at their mean values, see table~\ref{tab:astro}) to infer the joint posterior pdf of the DM mass and cross-section---as well as the nuisance parameters---preferred by the CoGeNT total event rate. The DM mass is sampled using a uniform prior in the range $1 \to 20$ GeV, while a uniform prior on log$(\sigma_n^{\rm SI}/{\rm cm^2})$ in the range $-42 \to -38$ is assumed for the spin-independent interaction. Along the lines of model 1b, we derive posterior pdfs for the fractional modulation amplitudes $S^i_{\rm m}$ via equation~(\ref{eq:permod}).   

\begin{figure}[t!]
\begin{minipage}[t]{0.49\textwidth}
\centering
\includegraphics[width=1.\columnwidth,trim=0mm 15mm 10mm 5mm, clip]{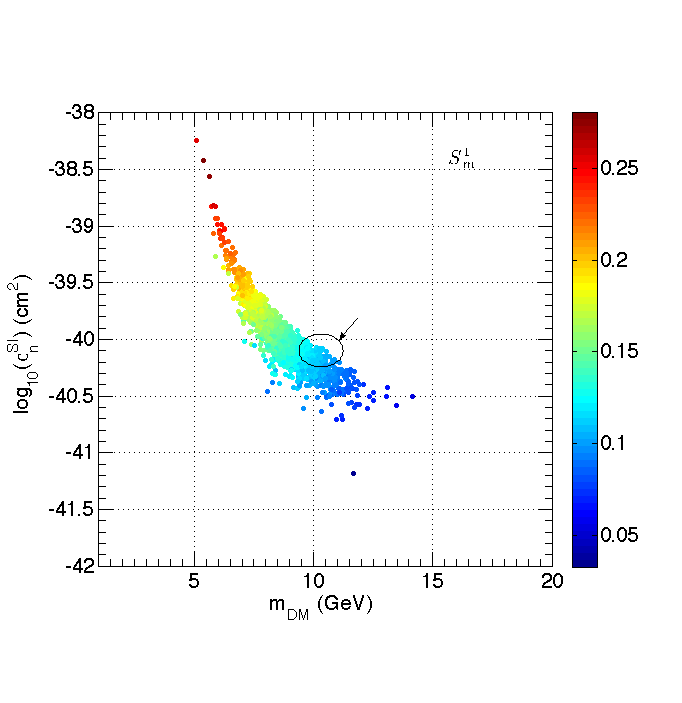}
\end{minipage}
\begin{minipage}[t]{0.49\textwidth}
\centering
\includegraphics[width=1.\columnwidth,trim=0mm 17mm 10mm 5mm, clip]{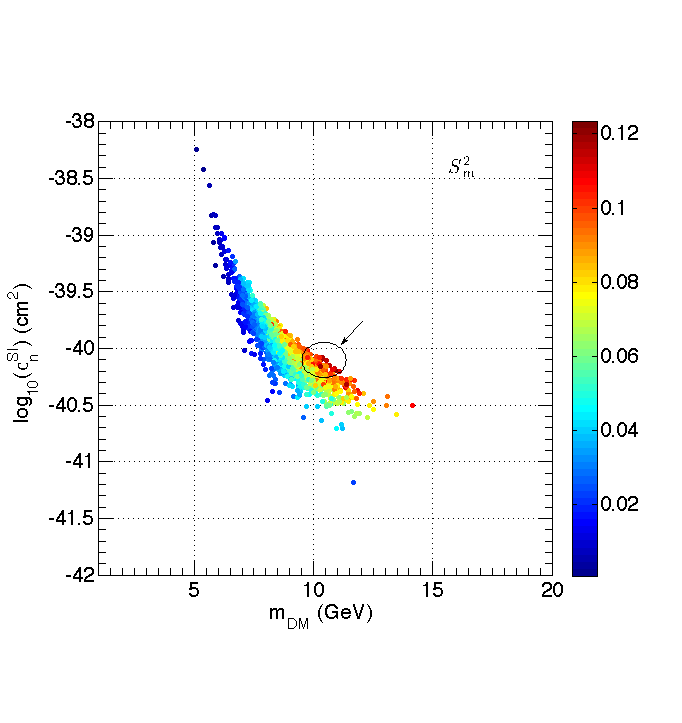}
\end{minipage}
\caption{3D marginal posteriors for $\{m_{\rm DM}, \sigma_n^{\rm SI},S_{\rm m\%}^i\}$, where the $S_{\rm m}^i$ direction is indicated by the colour coding,
for a specific realisation of an anisotropic DM velocity distribution (see main text for details).
The left panel shows $S_{\rm m}^1$,  and the right $S_{\rm m}^2$. The arrow in each plot points to a circled region in which the predicted values of
$S_{\rm m}^1$ and $S_{\rm m}^2$  find good agreement with the CoGeNT modulation data.
\label{fig:SmCogAnis}}
\end{figure}

Figure~\ref{fig:SmCogAnis} shows the resulting 3D marginal posteriors for $\{m_{\rm DM},\sigma_n^{\rm SI}, S_{\rm m}^i\}$ for the first two energy bins, $\Delta E_1$ and $\Delta E_2$.   Clearly, the predicted fractional modulation amplitudes in energy bins~1 and~2 have increased has a result of the anisotropic velocity distribution.
In bin~1, the modulation amplitude preferred by the total rate is $S_{\rm m}^1 = 0.14 \pm 0.03$ (compared with $S_{\rm m}^1 = 0.098 \pm 0.021$ for 
the isotropic DM model~1b), slightly larger than the $\sim 10 \to 11\%$ modulation favoured by the CoGeNT time-dependent data (see figure~\ref{fig:Smpdf}).
For energy bin~2, we find $S_{\rm m}^2 = 0.05 \pm 0.02$, which is an improvement on $S_{\rm m}^2 = 0.026 \pm 0.011$ predicted by model~1b, but still significantly lower than the 20\% required to explain the CoGeNT time modulation. In bin~3,  $S^3_{\rm m} = (0.12 \pm 0.77) \times 10^{-3}$  is consistent with no modulation and also with the expectations of models~1a and~1b.

Interestingly, although the ``best-fit'' $S_{\rm m}^1$ and especially $S_{\rm m}^2$ values are not perfect matches to the fractional modulation amplitudes required to fit the CoGeNT modulation data, we see from the 3D marginal posterior for $\{m_{\rm DM},\sigma_n^{\rm SI}, S_{\rm m}^i\}$ in figure~\ref{fig:SmCogAnis} that 
it is possible to find points in the preferred region in the DM parameter space that could improve further the agreement between the CoGeNT total and modulated rates. Take as an example  $m_{\rm DM} \simeq 10$ GeV and $\sigma_n^{\rm SI} \simeq 10^{-40}\  {\rm cm^2}$. These numbers lead to $S_{\rm m}^1 \simeq 0.10 \to 0.13$ and $S_{\rm m}^2 \simeq 0.12 \to 0.15$ (see the circled region in figure~\ref{fig:SmCogAnis}).
From this, we conclude that  an anisotropic DM velocity distribution appears to be a promising direction towards reducing the tension between the 
CoGeNT total and modulated rates. We defer a more complete exploration of this avenue to a future work.

\section{Conclusions\label{sec:concl}}

The  annual modulation signal claimed by the CoGeNT collaboration is an intriguing puzzle for the DM direct detection community. In this paper we have 
investigated the origin of the time-dependent signal detected by CoGeNT, and performed a sophisticated model comparison analysis to identify the best physical explanation. Our phenomenological approach features three nested scenarios: no modulation (amplitude of modulation set to zero), modulation due to DM (phase, modulation amplitude and period predicted by WIMP models), and modulation due to other physics (amplitude, phase and period of modulation as free parameters). We have used the Bayesian model comparison framework to assess the strength of evidence in favour of these scenarios. In addition, we have recomputed classical p-values against the null hypothesis of no modulation where analytically possible, since they had been misestimated in previous works.

Regarding the evidence for an annual modulation with a maximum rate occurring close to June~2 (or  $t_{\rm max} \sim  152$~days), we find that there is weak evidence for a modulation in both the energy ranges $0.5 \to 0.9$~keVee and $0.9 \to 3.0$~keVee.
This conclusion is borne out both from a Bayesian model comparison point of view and from a classical hypothesis testing perspective. The energy range $3.0 \to 4.5$~keVee shows no sign of modulation with a period of one year.  Modulation models due to ``other physics''  compare unfavourably with the no modulation case,  paying the price for their excessive complexity.

The question of whether this signal is due to DM or to other physical processes is more difficult to assess.
When all energy bins ($0.5 \to 4.5$~keVee) are considered, the DM models are strongly favoured over the phenomenological models in which the signal is due to other physics, with odds of several hundreds to one, depending on the details.  This conclusion is fairly robust with respect to changes in the priors assigned to the parameters of  the ``other physics'' models, in the sense that reasonable changes to the prior ranges could at most reduce the preference but not overturn the ranking.
 Interestingly, between the ``na\"{\i}ve" DM model~1a and the ``consistent'' model~1b which enforces self-consistency between the modulation signal and the CoGeNT total rate, there is no compelling preference for the latter, despite the fact that compatibility with the total rate dictates that the modulation amplitude be significantly smaller than the observed amplitude in the energy range $0.9 \to 3.0$~keVee. Nonetheless, we have shown that this tension between the total and the modulated rates could be partially remedied by using a triaxial DM velocity distribution with a tangential anisotropy.

\ack
We thank the CoGeNT collaboration for making the data publicly available.  CA thanks C.~Kelso for useful discussions and acknowledges use of the cosmo computing resources at CP3 of Louvain University. RT would like to thank David van Dyk for useful discussions. We thank T.~Schwetz and an anonymous referee for alerting us to a numerical error in an earlier version of the manuscript.

\appendix

\section*{References}

\bibliographystyle{iopart-num.bst}

\bibliography{biblio}

\providecommand{\newblock}{}
\begin{thebibliography}{10}
\expandafter\ifx\csname url\endcsname\relax
  \def\url#1{{\tt #1}}\fi
\expandafter\ifx\csname urlprefix\endcsname\relax\def\urlprefix{URL }\fi
\providecommand{\eprint}[2][]{\url{#2}}

\bibitem{Drukier:1986tm}
Drukier A, Freese K and Spergel D 1986 {\em Phys.Rev.\/} {\bf D33} 3495--3508

\bibitem{Freese:1987wu}
Freese K, Frieman J~A and Gould A 1988 {\em Phys. Rev.\/} {\bf D37} 3388

\bibitem{Bernabei:2008yi}
Bernabei R {\em et~al.\/} (DAMA) 2008 {\em Eur. Phys. J.\/} {\bf C56} 333--355
  (\textit{Preprint} \eprint{arXiv:0804.2741})

\bibitem{Bernabei:2010mq}
Bernabei R, Belli P, Cappella F, Cerulli R, Dai C {\em et~al.\/} 2010 {\em
  Eur.Phys.J.\/} {\bf C67} 39--49 (\textit{Preprint} \eprint{1002.1028})

\bibitem{Aalseth:2010vx}
Aalseth C {\em et~al.\/} (CoGeNT collaboration) 2011 {\em Phys.Rev.Lett.\/}
  {\bf 106} 131301 (\textit{Preprint} \eprint{1002.4703})

\bibitem{Angloher:2011uu}
Angloher G, Bauer M, Bavykina I, Bento A, Bucci C {\em et~al.\/} 2011
  (\textit{Preprint} \eprint{1109.0702})

\bibitem{Angle:2011th}
Angle J {\em et~al.\/} (XENON10 Collaboration) 2011 {\em Phys.Rev.Lett.\/} {\bf
  107} 051301 (\textit{Preprint} \eprint{1104.3088})

\bibitem{Chang:2010yk}
Chang S, Liu J, Pierce A, Weiner N and Yavin I 2010 {\em JCAP\/} {\bf 1008} 018
  (\textit{Preprint} \eprint{1004.0697})

\bibitem{Kopp:2011yr}
Kopp J, Schwetz T and Zupan J 2011  (\textit{Preprint} \eprint{1110.2721})

\bibitem{Kelso:2011gd}
Kelso C, Hooper D and Buckley M~R 2011  (\textit{Preprint} \eprint{1110.5338})

\bibitem{Aalseth:2011wp}
Aalseth C, Barbeau P, Colaresi J, Collar J, Diaz~Leon J {\em et~al.\/} 2011
  {\em Phys.Rev.Lett.\/} {\bf 107} 141301 (\textit{Preprint}
  \eprint{1106.0650})

\bibitem{Frandsen:2011ts}
Frandsen M~T, Kahlhoefer F, March-Russell J, McCabe C, McCullough M {\em
  et~al.\/} 2011 {\em Phys.Rev.\/} {\bf D84} 041301 (\textit{Preprint}
  \eprint{1105.3734})

\bibitem{Belli:2011kw}
Belli P, Bernabei R, Bottino A, Cappella F, Cerulli R {\em et~al.\/} 2011 {\em
  Phys.Rev.\/} {\bf D84} 055014 (\textit{Preprint} \eprint{1106.4667})

\bibitem{Cumberbatch:2011jp}
Cumberbatch D~T, L\'opez-Fogliani D~E, Roszkowski L, Ruiz~de Austri R and Tsai
  Y~L~S 2011  (\textit{Preprint} \eprint{1107.1604})

\bibitem{Farina:2011pw}
Farina M, Pappadopulo D, Strumia A and Volansky T 2011 {\em JCAP\/} {\bf 1111}
  010 (\textit{Preprint} \eprint{1107.0715})

\bibitem{Schwetz:2011xm}
Schwetz T and Zupan J 2011 {\em JCAP\/} {\bf 1108} 008 (\textit{Preprint}
  \eprint{1106.6241})

\bibitem{Fox:2011px}
Fox P~J, Kopp J, Lisanti M and Weiner N 2011  (\textit{Preprint}
  \eprint{1107.0717})

\bibitem{Hooper:2011hd}
Hooper D and Kelso C 2011 {\em Phys.Rev.\/} {\bf D84} 083001 (\textit{Preprint}
  \eprint{1106.1066})

\bibitem{Gao:2011ka}
Gao X, Kang Z and Li T 2011  (\textit{Preprint} \eprint{1107.3529})

\bibitem{Rajaraman:2011wf}
Rajaraman A, Shepherd W, Tait T~M and Wijangco A~M 2011  (\textit{Preprint}
  \eprint{1108.1196})

\bibitem{An:2011ck}
An H and Gao F 2011  (\textit{Preprint} \eprint{1108.3943})

\bibitem{DelNobile:2011je}
Del~Nobile E, Kouvaris C and Sannino F 2011 {\em Phys.Rev.\/} {\bf D84} 027301
  (\textit{Preprint} \eprint{1105.5431})

\bibitem{Foot:2011pi}
Foot R 2011 {\em Phys.Lett.\/} {\bf B703} 7--13 (\textit{Preprint}
  \eprint{1106.2688})

\bibitem{Boucenna:2011hy}
Boucenna M and Profumo S 2011 {\em Phys.Rev.\/} {\bf D84} 055011
  (\textit{Preprint} \eprint{1106.3368})

\bibitem{Frandsen:2011cg}
Frandsen M~T, Kahlhoefer F, Sarkar S and Schmidt-Hoberg K 2011 {\em JHEP\/}
  {\bf 1109} 128 (\textit{Preprint} \eprint{1107.2118})

\bibitem{Cerdeno:2011qv}
Cerdeno D~G, Huh J~H, Peiro M and Seto O 2011 {\em JCAP\/} {\bf 1111} 027
  (\textit{Preprint} \eprint{1108.0978})

\bibitem{Cline:2011zr}
Cline J~M and Frey A~R 2011 {\em Phys.Rev.\/} {\bf D84} 075003
  (\textit{Preprint} \eprint{1108.1391})

\bibitem{Fornengo:2011sz}
Fornengo N, Panci P and Regis M 2011 {\em Phys.Rev.\/} {\bf D84} 115002
  (\textit{Preprint} \eprint{1108.4661})

\bibitem{Kajiyama:2011fx}
Kajiyama Y, Okada H and Toma T 2011  (\textit{Preprint} \eprint{1109.2722})

\bibitem{McCabe:2011sr}
McCabe C 2011 {\em Phys.Rev.\/} {\bf D84} 043525 (\textit{Preprint}
  \eprint{1107.0741})

\bibitem{Natarajan:2011gz}
Natarajan A, Savage C and Freese K 2011 {\em Phys.Rev.\/} {\bf D84} 103005
  (\textit{Preprint} \eprint{1109.0014})

\bibitem{Frandsen:2011gi}
Frandsen M~T, Kahlh{\"o}fer F, McCabe C, Sarkar S and Schmidt-Hoberg K 2011
  (\textit{Preprint} \eprint{1111.0292})

\bibitem{Trotta:2005ar}
Trotta R 2007 {\em Mon.Not.Roy.Astron.Soc.\/} {\bf 378} 72--82
  (\textit{Preprint} \eprint{astro-ph/0504022})

\bibitem{Kunz:2006mc}
Kunz M, Trotta R and Parkinson D 2006 {\em Phys.Rev.\/} {\bf D74} 023503
  (\textit{Preprint} \eprint{astro-ph/0602378})

\bibitem{Trotta:2008qt}
Trotta R 2008 {\em Contemp.Phys.\/} {\bf 49} 71--104 (\textit{Preprint}
  \eprint{0803.4089})

\bibitem{Sellke:2001}
Sellke T, Bayarri M~J and Berger J~O 2001  {\bf 55} 62--71

\bibitem{Vardanyan:2011in}
Vardanyan M, Trotta R and Silk J 2011  (\textit{Preprint} \eprint{1101.5476})

\bibitem{Gordon:2007xm}
Gordon C and Trotta R 2007 {\em Mon.Not.Roy.Astron.Soc.\/} {\bf 382} 1859--1863
  (\textit{Preprint} \eprint{0706.3014})

\bibitem{Martin:2010hh}
Martin J, Ringeval C and Trotta R 2011 {\em Phys.Rev.\/} {\bf D83} 063524
  (\textit{Preprint} \eprint{1009.4157})

\bibitem{Sapone:2010uy}
Sapone D, Kunz M and Amendola L 2010 {\em Phys.Rev.\/} {\bf D82} 103535
  (\textit{Preprint} \eprint{1007.2188})

\bibitem{March:2010ex}
March M, Starkman G, Trotta R and Vaudrevange P 2011 {\em
  Mon.Not.Roy.Astron.Soc.\/} {\bf 410} 2488--2496 (\textit{Preprint}
  \eprint{1005.3655})

\bibitem{Ichikawa:2009ir}
Ichikawa K, Kawasaki M, Nakayama K, Sekiguchi T and Takahashi T 2009 {\em
  JCAP\/} {\bf 0908} 013

\bibitem{Feroz:2009dv}
Feroz F, Hobson M~P, Roszkowski L, Ruiz~de Austri R and Trotta R 2009
  (\textit{Preprint} \eprint{0903.2487})

\bibitem{Cabrera:2010xx}
Cabrera M~E, Casas J, Ruiz~de Austri R and Trotta R 2011 {\em Phys.Rev.\/} {\bf
  D84} 015006 (\textit{Preprint} \eprint{1011.5935})

\bibitem{Arina:2011xu}
Arina C 2011  (\textit{Preprint} \eprint{1110.0313})

\bibitem{Feroz:2007kg}
Feroz F and Hobson M 2008 {\em Mon.Not.Roy.Astron.Soc.\/} {\bf 384} 449
  (\textit{Preprint} \eprint{0704.3704})

\bibitem{Feroz:2008xx}
Feroz F, Hobson M and Bridges M 2009 {\em Mon.Not.Roy.Astron.Soc.\/} {\bf 398}
  1601--1614 (\textit{Preprint} \eprint{0809.3437})

\bibitem{collar}
Collar J {\em How to use time-stamped CoGeNT data.\/}

\bibitem{Aalseth:2008rx}
Aalseth C {\em et~al.\/} (CoGeNT Collaboration) 2008 {\em Phys.Rev.Lett.\/}
  {\bf 101} 251301 (\textit{Preprint} \eprint{0807.0879})

\bibitem{Arina:2011si}
Arina C, Hamann J and Wong Y~Y~Y 2011 {\em JCAP\/} {\bf 1109} 022
  (\textit{Preprint} \eprint{1105.5121})

\bibitem{Green:2003yh}
Green A~M 2003 {\em Phys. Rev.\/} {\bf D68} 023004 (\textit{Preprint}
  \eprint{astro-ph/0304446})

\bibitem{Fornengo:2003fm}
Fornengo N and Scopel S 2003 {\em Phys.Lett.\/} {\bf B576} 189--194
  (\textit{Preprint} \eprint{hep-ph/0301132})

\bibitem{Freese:2003na}
Freese K, Gondolo P, Newberg H~J and Lewis M 2004 {\em Phys.Rev.Lett.\/} {\bf
  92} 111301 (\textit{Preprint} \eprint{astro-ph/0310334})

\bibitem{Savage:2006qr}
Savage C, Freese K and Gondolo P 2006 {\em Phys.Rev.\/} {\bf D74} 043531
  (\textit{Preprint} \eprint{astro-ph/0607121})

\bibitem{Green:2010gw}
Green A~M 2010 {\em JCAP\/} {\bf 1010} 034 (\textit{Preprint}
  \eprint{1009.0916})

\bibitem{Kelso:2010sj}
Kelso C and Hooper D 2011 {\em JCAP\/} {\bf 1102} 002 (\textit{Preprint}
  \eprint{1011.3076})

\bibitem{Navarro:1996gj}
Navarro J~F, Frenk C~S and White S~D~M 1997 {\em Astrophys. J.\/} {\bf 490}
  493--508 (\textit{Preprint} \eprint{astro-ph/9611107})

\bibitem{Salucci:2007tm}
Salucci P, Lapi A, Tonini C, Gentile G, Yegorova I {\em et~al.\/} 2007 {\em
  Mon.Not.Roy.Astron.Soc.\/} {\bf 378} 41--47 (\textit{Preprint}
  \eprint{astro-ph/0703115})

\bibitem{Donato:2009ab}
Donato F, Gentile G, Salucci P, Martins C, Wilkinson M {\em et~al.\/} 2009

\bibitem{Reid:2009nj}
Reid M, Menten K, Zheng X, Brunthaler A, Moscadelli L {\em et~al.\/} 2009 {\em
  Astrophys.J.\/} {\bf 700} 137--148 (\textit{Preprint} \eprint{0902.3913})

\bibitem{Gillessen:2008qv}
Gillessen S, Eisenhauer F, Trippe S, Alexander T, Genzel R {\em et~al.\/} 2009
  {\em Astrophys.J.\/} {\bf 692} 1075--1109 (\textit{Preprint}
  \eprint{0810.4674})

\bibitem{Smith:2006ym}
Smith M~C, Ruchti G, Helmi A, Wyse R, Fulbright J {\em et~al.\/} 2007 {\em
  Mon.Not.Roy.Astron.Soc.\/} {\bf 379} 755--772 (\textit{Preprint}
  \eprint{astro-ph/0611671})

\bibitem{Dehnen:1997cq}
Dehnen W and Binney J 1998 {\em Mon. Not. Roy. Astron. Soc.\/} {\bf 298}
  387--394 (\textit{Preprint} \eprint{astro-ph/9710077})

\bibitem{Weber:2009pt}
Weber M and de~Boer W 2010 {\em Astron.Astrophys.\/} {\bf 509} A25
  (\textit{Preprint} \eprint{0910.4272})

\bibitem{Salucci:2010qr}
Salucci P, Nesti F, Gentile G and Martins C 2010 {\em Astron.Astrophys.\/} {\bf
  523} A83 (\textit{Preprint} \eprint{1003.3101})

\bibitem{Dehnen:2006cm}
Dehnen W, McLaughlin D and Sachania J 2006 {\em Mon.Not.Roy.Astron.Soc.\/} {\bf
  369} 1688--1692 (\textit{Preprint} \eprint{astro-ph/0603825})

\bibitem{Sakamoto:2002zr}
Sakamoto T, Chiba M and Beers T~C 2003 {\em Astron.Astrophys.\/} {\bf 397}
  899--912 (\textit{Preprint} \eprint{astro-ph/0210508})

\bibitem{Navarro:2008kc}
Navarro J~F {\em et~al.\/} 2008  (\textit{Preprint} \eprint{0810.1522})

\bibitem{Wilks:1938}
Wilks S~S 1938 {\em The Annals of Mathematical Statistics\/}

\bibitem{Protassov:2002sz}
Protassov R, van Dyk D~A, Connors A, Kashyap V~L and Siemiginowska A 2002 {\em
  The Astrophysical Journal\/} {\bf 571}

\bibitem{Chernoff:1954}
Chernoff H 1954 {\em The Annals of Mathematical Statistics\/}

\bibitem{Shapiro:1988}
Shapiro A 1988 {\em International Statistical Review\/} {\bf 56} 49--62

\bibitem{Evans:2000gr}
Evans N, Carollo C and de~Zeeuw P 2000 {\em Mon.Not.Roy.Astron.Soc.\/} {\bf
  318} 1131 (\textit{Preprint} \eprint{astro-ph/0008156})

\bibitem{Belli:2002yt}
Belli P, Cerulli R, Fornengo N and Scopel S 2002 {\em Phys.Rev.\/} {\bf D66}
  043503 (\textit{Preprint} \eprint{hep-ph/0203242})

\end{thebibliography}

\end{document}